\documentclass[a4paper]{report}
\usepackage[utf8]{inputenc}
\usepackage[T1]{fontenc}
\usepackage{RJournal}
\usepackage{amsmath,amssymb,array}
\usepackage{booktabs}
\usepackage{textcomp}
\usepackage{bbm}

\begin{document}

\sectionhead{}

\begin{article}

\title{\pkg{SSNbayes}: An R package for Bayesian spatio-temporal modelling on stream networks}
\author{by Edgar Santos-Fernandez, Jay M. Ver Hoef, James M. McGree, Daniel J. Isaak, Kerrie Mengersen and Erin E. Peterson}
\maketitle

\abstract{
Spatio-temporal models are widely used in many research areas from ecology to epidemiology. However, most covariance functions describe spatial relationships based on Euclidean distance only. 
In this paper, we introduce the \texttt{R} package \pkg{SSNbayes} for fitting Bayesian spatio-temporal models and making predictions on branching stream networks. 
\pkg{SSNbayes} provides a linear regression framework with multiple options for incorporating spatial and temporal autocorrelation. 
Spatial dependence is captured using stream distance and flow connectivity while temporal autocorrelation is modelled using vector autoregression approaches.  
\pkg{SSNbayes} provides the functionality to make predictions across the whole network, compute exceedance probabilities and other probabilistic estimates such as the proportion of suitable habitat.
We illustrate the functionality of the package using a stream temperature dataset collected in Idaho, USA. }

\section{Introduction}

Rivers and streams are of vital ecological and economic importance \citep{vorosmarty2010global} but are under pressure from anthropogenic impacts such as climate change, pollution, water extractions and overfishing. 
In the past, data describing critical characteristics such as nutrients, sediments, pollutants and stream flow tended to be sparse in space and/or time. However, recent developments in in-situ sensor technology are revolutionizing ecological research and natural resource monitoring. These new data sets create exciting opportunities to measure, learn about, and manage spatio-temporal dynamics of stream attributes. A number of free software packages for stream network modelling have been developed in the literature \citep{hoef2014ssn, rtop, smnet}, which account for the unique spatial relationships found in streams data (e.g. network structure, longitudinal (upstream/downstream) connectivity, water flow volume and direction). However, they are not designed to simultaneously account for the temporal variability that often accompanies spatial variation in the new data sets derived from modern sensor arrays. For example, the package \textsf{SSN} \citep{hoef2014ssn} fits spatial regression models for stream networks. Currently, changes over time can only be incorporated using random effects. Similarly, the additive models that can be fitted using the package \textsf{smnet} \citep{smnet} are purely spatial.

There are several \textsf{R} packages for spatio-temporal modelling that are described in the Space-time CRAN Task View \citep{cranSpaceTime}. For example, spatial/temporal dependence can be incorporated via the nlme package \textsf{nlme} package \citep{nlme} and other packages such as spBayes \textsf{spBayes} \citep{spBayes} allow random effects modelling for point-referenced data. 
The well-known package \textsf{CARBayes} \citep{CARBayes} contains useful tools for implementing Bayesian spatial models using random effects via conditional autoregressive (CAR) priors. 
The package \textsf{RandomFields}  \citep{RandomFields} also allows the generation of spatial process data based on multiple kernels. 
Similarly, \textsf{geoR} \citep{geoR} contain useful tools for spatial kriging and interpolation. One the most popular implementations among practitioners is the \textsf{R-INLA} package \citep{rinla}, which uses approximate Bayesian inference to include multiple spatiotemporal modelling options e.g. stochastic partial differential equation (SPDE) - autoregressive (AR) models. \textsf{FRK} \citep{FRK} harnesses the use of spatial basis functions and discrete areal units with a focus on large datasets but none of these packages are designed for networked systems. 

Here, we describe the \pkg{SSNbayes} package that has been designed to address many of the limitations of current software tools for spatio-temporal modelling on stream networks. This package is equipped to fit spatio-temporal stream network models and produce predictions in space and time incorporating uncertainty.
It uses the Bayesian inference machinery and particularly the probabilistic programming language \pkg{Stan} \citep{carpenter2017stan}.
 
In the next section (Methods) we introduce the relevant statistical models and follow that with a software application to a stream temperature data set before a discussion section.

\section{Methods}
\label{sec:met}

Consider the following spatio-temporal linear model:

\begin{equation}
\pmb{y} = \pmb{X}\pmb{\beta} + \pmb{v} + \pmb{\epsilon},
\label{eq:lm0}
\end{equation}

\noindent where the response variable $\pmb{y} = [\pmb{y}_1, \pmb{y}_2,\cdots,\pmb{y}_T]$ is a stacked vector of length $n = S\times T$ for $S$ spatial locations and $T$ time points. 
The vector $\pmb{y}_1$ contains the observations at the spatial locations $S$ at time $t=1$.
Let $\pmb{X}$ be a $n \times p$ design matrix of $p$ covariates, $\pmb{X} = [\pmb{X}_1, \pmb{X}_2,\cdots,\pmb{X}_T]$ and $\pmb{\beta}$ a $p \times 1$ vector of regression coefficients.
$\pmb{v}$ is a vector of length $n$ spatial autocorrelated random effects, which can be modelled, for instance, using a Gaussian process \citep{banerjee2014hierarchical}.
The vector $\pmb{\epsilon}$ is the independent unstructured random error term where $\textrm{var}(\pmb{\epsilon}) = \sigma^2_0\pmb{I}$. The parameter $\sigma_{0}^2$ is called the nugget effect and $\pmb{I}$ is the identity matrix.

\begin{figure}[htbp]
  \centering
   \includegraphics[width=5.25in]{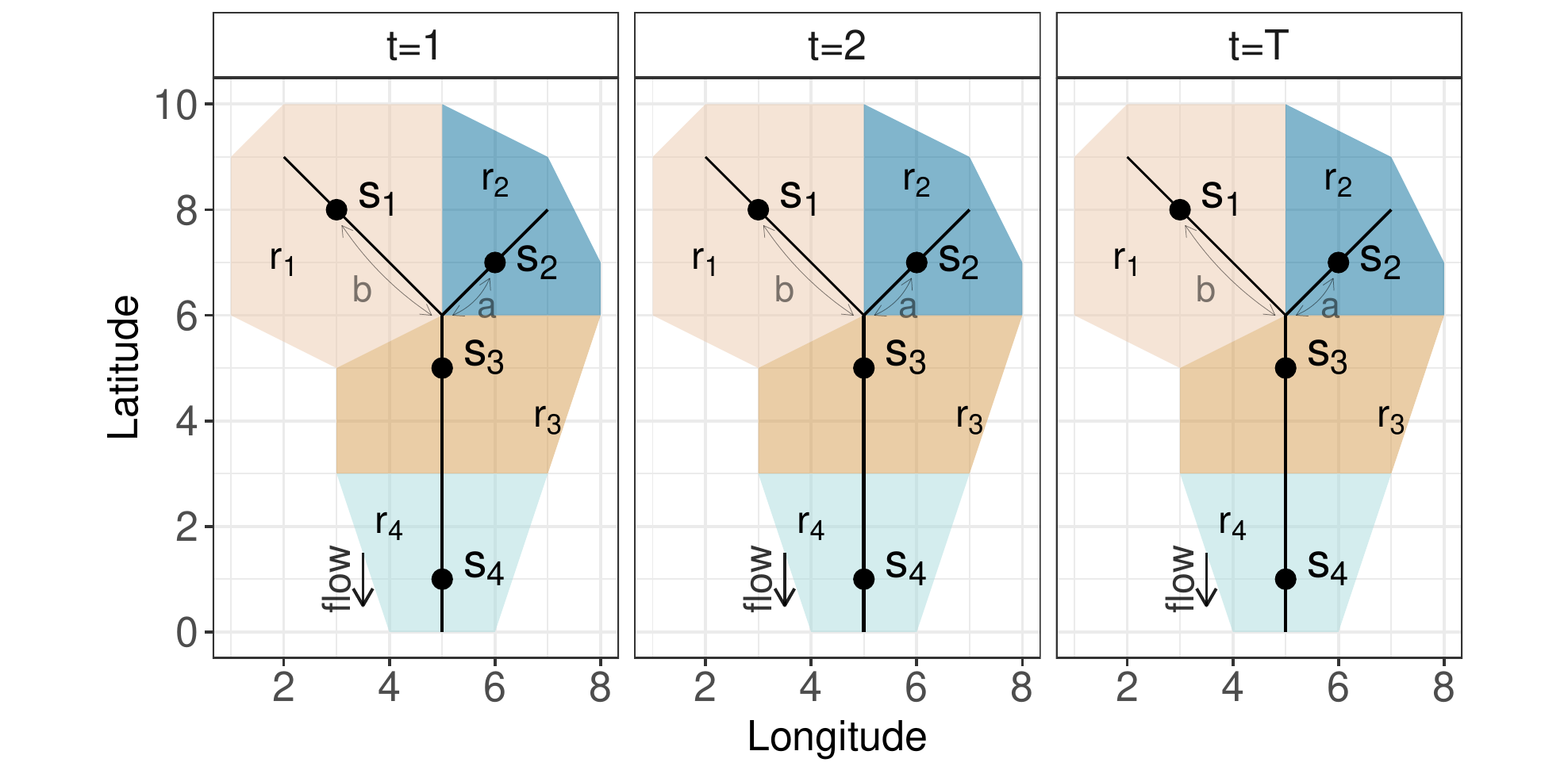}
  \caption{Stream network across multiple time points. Four spatial locations ($s_1-s_4$) and four regions ($r_1-r_4)$ are depicted. }
  \label{fig:network}
\end{figure}

The network shown in Fig.\ref{fig:network} 
represents repeated measures at several time points $t$ from four spatial locations ($s_1$ to $s_4$). 
The direction of the water flow is also shown in the figure (from North to South) and the stream outlet (most downstream point) is below the location $s_4$. 
Spatial locations $s_1$ and $s_3$ share water flow (flow-connected) while $s_1$ and $s_2$ are not (flow-unconnected).
In this figure we identify a confluence as the junction between the segments where $s_1$ and $s_2$ are located. 
The distances to the junction are $a$ and $b$, where $a \leqslant b$.

\subsection*{Spatial stream network models}

In this section, we describe the purely spatial models that arise when we consider the model from Eq~\ref{eq:lm0} at a unique time point. 

There are several ways of incorporating spatial autocorrelation.  
Multiple covariance models have been proposed specifically for stream networks to capture particular types of spatial dependence related to network structure and stream flow \citep{ver2006spatial, cressie2006spatial}.

Fundamental to modelling spatial dependence in stream networks is capturing which sites are flow connected, and thus potentially have dependent data due to shared water flow.  
\citet{ver2006spatial} proposed covariance functions based on flow connectivity and spatial weights known as tail-up models that can capture spatial dependence as a function of stream distance.
These methods were extended by \citet{ver2010moving} to include a family of tail-down models that can also explain dependence between locations that are not connected by flow.

These covariance matrices have been used in several subsequent applications. For example, \citet{money2009using, money2009modern} used them to construct space-time covariance models. 
Other examples can be found in  \citet{isaak2014applications, mcmanus2020variation, jackson2018spatio, rodriguez2019spatial}

\noindent 
\subsection*{Euclidean distance models}

A typical modelling approach is to capture spatial dependence in  $\pmb{v}$ from Eq~\ref{eq:lm0} via the second moment, in which the amount of autocorrelation decays with the Euclidean distance. 
Some of the most common covariance functions are the exponential, Gaussian and spherical \citep{cressie2015statistics, banerjee2014hierarchical}: 

\begin{equation}
    \textrm{exponential model,} \ \  C_{ED}(d\mid \pmb{\theta}) = \sigma_e^2 e^{-3d/\alpha_e}, \alpha_e \in (0,\infty), \sigma_e ^ 2 > 0,
    \label{eq:ced}
\end{equation}
\begin{equation}
    \textrm{Gaussian model,} \ \ C_{ED}(d\mid \pmb{\theta}) = \sigma_e^2 e^{-3(d/\alpha_e)^2},
\label{eq:ced2}
\end{equation}
and
\begin{equation}
\textrm{spherical model,} \ \ C_{ED}(d\mid \pmb{\theta}) = \sigma_e^2\left(1-\frac{3d}{2\alpha_e} + \frac{d^3}{2\alpha_e^3}\right)\mathbbm{1}(d/\alpha_e \leqslant 1),
\label{eq:ced3}
\end{equation}

\noindent where $d$ is the Euclidean distance between two locations $s_i$ and $s_j$.
$\pmb{\theta}$ represents the spatial parameters ($\alpha_e$, $\sigma_e^2$), where 
$\sigma_e^2$ is the partial sill, $\alpha_e$ is the spatial range parameter and $\mathbbm{1}(\cdot)$ is the indicator function. 
The partial sill is the variance between two uncorrelated locations. Negligible spatial correlation is assumed between points located at a distance greater than the spatial range parameter.

\subsection{Tail-up models}
Tail-up models in stream networks were developed by convolving a moving average function with white noise \citep{ver2006spatial}. As the name suggests, the moving average function points in the upstream direction from a stream location in a tail-up model. This restricts correlation to flow-connected locations. In addition, the function must be split upstream at junctions and spatial weights are included to maintain stationary variances by controlling the proportion allocated to each upstream segment.

Given a pair of sites $s_i$ and $s_j$, the tail-up covariance matrix is defined as:

$$C_{TU}(s_i,s_j|\pmb{\theta}) = \left\{\begin{matrix}
0 \:\:\: \textrm{if $s_i,s_j$ are flow-unconnected}, \\C_u(h\mid \pmb{\theta}) W_{ij} \:\:\: \textrm{if $s_i,s_j$ are flow-connected}, \end{matrix}\right.$$

\noindent where $C_u(h\mid \pmb{\theta})$ is an unweighted tail-up covariance between two locations and the $W_{ij}$ represents the spatial weights between sites $i$ and $j$ and is defined by the branching structure of the network, the watershed area or other spatial variable used. 
Let $h$ be the hydrologic distance between sites, then $C_u$ can take a variety of forms including:

\begin{align}
\textrm{Tail-up exponential model},& \ \ C_{u}(h \mid \pmb{\theta}) = \sigma_u^2 e^{-3h/\alpha_u}, \\
\textrm{Tail-up linear-with-sill model},& \ \ C_u(h \mid \pmb{\theta}) = \sigma_u^2 (1-h/\alpha_u)\mathbbm{1}(h/\alpha_u \leqslant 1), \\
\textrm{Tail-up spherical model},& \ \ C_u(h \mid \pmb{\theta}) = \sigma_u^2 \left(1-\frac{3h}{2\alpha_u} + \frac{h^3}{2\alpha_u^3}\right)\mathbbm{1}(h/\alpha_u \leqslant 1)
\end{align}

\noindent where $\sigma^2_{u}$ is the partial sill and $\alpha_u$ is the range parameter.

\subsection{Tail-down models}
Tail-down models were developed by convolving a moving average function with white noise strictly downstream from a stream location \citep{ver2010moving}. Tail-down models differ from tail-up models because they allow spatial correlation between both flow-connected and flow-unconnected locations.

Consider two flow-unconnected sites (e.g. $s_1$ and $s_2$ in Fig.\ref{fig:network}). Define $a$ and $b$ as the hydrologic distance from $s_1$ and $s_2$ to their common confluence so that $a \leqslant b$. The tail-down models are defined as follows:

Tail-down exponential model,
$$C_{TD}(a,b, h|\pmb{\theta}) = \left\{\begin{matrix}
\sigma_d^2 e^{-3h/\alpha_d} \:\:\: \textrm{if flow-connected,} \\
\sigma_d^2 e^{-3(a+b)/\alpha_d} \:\:\: \textrm{if flow-unconnected,} \end{matrix}\right. $$
Tail-down linear-with-sill model,
$$C_{TD}(a,b, h|\pmb{\theta}) = \left\{\begin{matrix}
\sigma_d^2 (1-\frac{h}{\alpha_d})\mathbbm{1}(\frac{h}{\alpha_d} \leqslant 1) \:\:\: \textrm{if flow-connected,} \\
\sigma_d^2 (1-\frac{b}{\alpha_d})\mathbbm{1}(\frac{b}{\alpha_d} \leqslant 1) \:\:\: \textrm{if flow-unconnected,} \end{matrix}\right. $$
Tail-down spherical model,
$$C_{TD}(a,b, h|\pmb{\theta}) = \left\{\begin{matrix}
\sigma_d^2 (1-\frac{3h}{2\alpha_d} + \frac{h^3}{2\alpha_d^3})\mathbbm{1}(\frac{h}{\alpha_d} \leqslant 1) \:\:\: \textrm{if flow-connected,}\\
\sigma_d^2 (1-\frac{3a}{2\alpha_d} + \frac{b}{2\alpha_d})(1-\frac{b}{\alpha_d})\mathbbm{1}(\frac{b}{\alpha_d} \leqslant 1) \:\:\: \textrm{if flow-unconnected,} \end{matrix}\right. $$
where $\sigma_d^2$ is the partial sill, $\alpha_d$ is the range parameter, and $\mathbbm{1}(\cdot)$ is the indicator function, equal to 1 if its argument is true, otherwise it is zero.

Spatial dependence in stream networks is influenced by many factors such as climatic gradients,
passive movement of nutrients and sediments downstream, and the movement of organisms \citep{peterson2013modelling}.
To capture these complex spatial patterns a mixture of Euclidean, tail-up and tail-down covariance matrices is often used.

In Eq~\ref{eq:lm0}, for a purely spatial case $\pmb{v}$ is a vector of dimension $s$ corresponding to the spatial locations, with covariance matrix $\pmb{\Sigma} = COV(\pmb{v})$.

\begin{equation}
\pmb{\Sigma} = COV(\pmb{v}) = \pmb{C}_{ED} + \pmb{C}_{TU} + \pmb{C}_{TD}   
= \sigma_e^2\pmb{R}_{e}(\alpha_e)+\sigma_u^2\pmb{R}_{u}(\alpha_u) + \sigma_d^2\pmb{R}_{d}(\alpha_d), 
\label{eq:covs}
\end{equation}

\noindent where $\sigma^2_{e}$, $\sigma^2_{u}$, and $\sigma^2_{d}$ are the partial sills for Euclidean, tail-up and tail-down functions, respectively. The correlation matrices $\pmb{R}_{u}(\alpha_u)$, $\pmb{R}_{d}(\alpha_d)$ and $\pmb{R}_{e}(\alpha_e)$ are obtained as a function of the range parameters $\alpha_u$, $\alpha_d$ and $\alpha_e$ \citep{hoef2014ssn}.

For space-time applications, we can use the same spatial covariance matrix or we can build a dynamic model with spatial parameters that are time-specific. In this work, we opted for the first approach, since this reduces the number of parameters to be estimated from the model and is less computationally demanding. We return to this point in the Discussion.

\subsection{Spatio-temporal stream network models}

Following the above discussion, consider the stream network in Fig.\ref{fig:network}, that evolves across discrete time points $t = 1, 2, \ldots, T$.
Let a response variable $\pmb{y}_t$ be an $S \times 1$ vector of random variables at unique and fixed spatial locations of $s = 1, 2, \ldots, S$. 
We start with the following conditional spatio-temporal model:

\begin{equation}
[\pmb{y}_1,\pmb{y}_2,\cdots,\pmb{y}_T] =  \prod_{t=2}^{T}[\pmb{y}_t \mid  \pmb{y}_{t-1}, \pmb{\theta},\pmb{X}_{t},\pmb{X}_{t-1},\pmb{\beta}, \pmb{\Phi}_1, \pmb{\Sigma}][\pmb{y}_1] 
\label{eq:VAR1def}
\end{equation}
where $\pmb{y}_1$ is the process at $t=1$, and
\begin{equation}
[\pmb{y}_t \mid  \pmb{y}_{t-1},\pmb{\theta},\pmb{X}_{t},\pmb{X}_{t-1},\pmb{\beta}, \pmb{\Phi}_1, \pmb{\Sigma}  ] = \mathcal{N}(\pmb{\mu}_{t},\pmb{\Sigma} + \sigma^2_0\pmb{I}), 
\end{equation}
and the mean can be expressed as follows:
\begin{equation}
\pmb{\mu}_{t} = \pmb{X}_{t}\pmb{\beta} + \pmb{\Phi}_1 (\pmb{y}_{t-1} - \pmb{X}_{t-1}\pmb{\beta}),
\label{eq:err_}
\end{equation}

\noindent Here, $\pmb{\Sigma} = COV(\pmb{v})$ is an $S \times S$ spatial covariance matrix of the form given in Eq~\ref{eq:covs} and Eq~\ref{eq:err_} is a vector autoregressive process of order one VAR(1) \citep{hamilton1994time}. The square transition matrix, $\pmb{\Phi}_1$ of dimension $S \times S$, has elements $\phi_{ij}$, which describe the amount of temporal autocorrelation between two spatial locations $i$ and $j$.

\subsection {Vector autoregressive model variations}

\label{sec:dst}

Two variations of the vector autoregressive spatial process have been implemented in the \pkg{SSNbayes} package to incorporate temporal dependence.

\noindent {\bf Case 1 (AR)}

\noindent The simplest case considers the same temporal autocorrelation for all spatial locations. Therefore all the diagonal elements of $\pmb{\Phi}_1$ are equal to $\phi$ and all the off-diagonal elements are set to zero, which assumes negligible cross-correlation between time series. That is:

\begin{equation}\label{eq:case1}
\Phi_1 = \begin{bmatrix}
\phi & 0 & \cdots & 0 \\ 
0 & \phi &  \cdots  & 0 \\ 
\vdots & \vdots &  \ddots & \vdots \\ 
0 & 0  & \cdots & \phi \\
\end{bmatrix}.
\end{equation}

Spatial locations in large river networks often have different elevations, climatic conditions, or local flow regimes and this can affect the amount of temporal autocorrelation found in observations. Hence, the assumption that there is a common $\phi$ for all locations may not always be appropriate and this motivates Case 2.

\noindent{\bf Case 2 (VAR)}

\noindent The second method considers the parameter $\phi$ to be site specific ($\phi_{1}, \phi_{2}, \cdots, \phi_{S}$), which is known as the autoregressive shock model \citep{wikle1998hierarchical}, which can be defined through $\Phi_1$ as follows:

\begin{equation}\label{eq:case2}
\Phi_1 = \begin{bmatrix}
\phi_{1} & 0 & \cdots & 0 \\ 
0 & \phi_{2} &  \cdots  & 0 \\ 
\vdots & \vdots &  \ddots & \vdots \\ 
0 & 0  & \cdots & \phi_{S} \\ 
\end{bmatrix}.
\end{equation}

Other VAR structures consider $\phi$ as a linear combination of spatial covariates and cross-correlation between time series \citep{santos2021bayesian}. This variation is not currently implemented in \pkg{SSNbayes} but is under development.

\pkg{SSNbayes} uses Hamiltonian Monte Carlo (HMC) simulations via \pkg{rstan}\citep{carpenter2017stan}.
Formulating this model in a Bayesian framework requires sampling from the following posterior distribution:

\begin{equation}
[\pmb{\beta},  \pmb{\Phi}_{1}, \sigma^2_{0}, \sigma^2_{u}, \alpha_{u}, \sigma^2_{d}, \alpha_{d}, \sigma^2_{e}, \alpha_{e}\mid \mathbf{y}, \mathbf{X}].
\end{equation}

We also need to define prior distributions for the parameters of interest. Currently, non-informative prior distributions are the only option in \texttt{ssnbayes}, but the functionality to include other prior distributions may be included in future package versions. 
The implemented prior distributions are the following:

\begin{flalign*}
\phi &\sim \textrm{Uniform}\left(-1,1\right)  && \text{\# uniform prior on the autoregressive parameter}\\
\alpha_{.}&\sim \textrm{Uniform}\left(0, 4 \times \max(h) \right)   && \text{\# non-informative prior on the spatial range}\\
\sigma_{.}&\sim \textrm{Uniform}\left(0,100\right)   && \text{\# non-informative prior on the partial sill SD}\\
\sigma_{0}&\sim \textrm{Uniform}\left(0,100\right)   && \text{\# non-informative prior on the nugget effect SD}\\
\beta & \sim \mathcal{N}\left(0,1000\right)  && \text{\# prior on the regression coefficients (intercept and slope)}\\
\end{flalign*}

For the autoregressive parameter ($\phi$), a uniform prior defined from -1 to 1 is used to ensure the process is stationary. 
The upper limit for the spatial range parameter is set to four times the maximum hydrologic distance between observation locations on the network. 

There are two ways of making predictions in \pkg{SSNbayes}. By default, predictions are produced for missing values (NA) in the response variable in the observation dataset used to fit the model employing the posterior predictive distributions:

\begin{equation}
p(\hat{\mathbf{y}} \mid \mathbf{y}, \mathbf{X}, \hat{\mathbf{X}}) = 
\int p(\hat{\mathbf{y}} \mid \theta, \hat{\mathbf{X}}) p(\theta \mid \mathbf{y}, \mathbf{X}) \, \textrm{d}\theta.
    \label{eq:pred}
\end{equation}

However, this approach is not recommended when making predictions at a large number of spatial locations (e.g. predicting over an extensive branching stream network). 

The second approach uses the fitted model produced using \texttt{ssnbayes} to generate predictions using the simple kriging predictor in a prediction dataset. This produces estimates as a weighted average of observations:

\begin{equation}
\widehat{\pmb{y}}_{P} = \pmb{X}_{P}\pmb{\beta} + {\pmb{C}_{OP}}'\pmb{C}^{-1}_{OO} (\pmb{y}_{O} - \pmb{X}_{O}\pmb{\beta} ),\\
\label{eq:krig}
\end{equation}

\noindent where subscripts $O$ and $P$ indicate the observation and prediction locations, respectively. The stacked vector $\widehat{\pmb{y}}_{P}$ contains the predictions at the $P$ spatial locations  across all the time points $T$. The observations are represented in the stacked vector $\pmb{y}_{O}$, which contains all of the observations across the $T$ time points. 
$\pmb{X}_{P}$ and $\pmb{X}_{O}$ are space-time design matrices of covariates for the observations and predictions, respectively, while $\pmb{\beta}$ is a vector of regression coefficients.

The matrix $\pmb{C}_{OO}$ of dimension $O \times T$ by $O \times T$, contains the covariance between observations at all time points, where $O$ and $T$ are the number of observation and time points respectively. Similarly,  $\pmb{C}_{OP}$ is a $O \times T$ by $P \times T$ rectangular matrix of covariances between observation and prediction locations at all time points with the same structure as $\pmb{C}_{OO}$. That is,  $\pmb{C}_{OO}$ was obtained from an AR exponential tail-down model with parameters $\phi$, $\sigma_{td}$ and $\alpha_{td}$, these same parameters are used to construct $\pmb{C}_{OP}$.

The covariance matrix of observations ($\pmb{C}_{OO}$) must be inverted at each MCMC iteration when making predictions (Eq.~\ref{eq:krig}) and this quickly becomes computationally challenging for large datasets. However, using the Kronecker product significantly reduces the computational burden in these cases \citep{wikle2019spatio}: 
$$C^{-1}_{OO} = \pmb{\Sigma}_{OO}^{-1} \otimes \pmb{\Sigma}_{var}^{-1}$$
\noindent where $\pmb{\Sigma}_{OO}$ is the spatial covariance matrix defined in Eq~\ref{eq:covs} and $\pmb{\Sigma}_{var}$ is the temporal covariance matrix of the VAR(1) process.

\section{The \pkg{SSNbayes} package}
\label{sec:ssn}

\subsection{Installation}
The \pkg{SSNbayes} package can be found in CRAN and Github (\url{https://github.com/EdgarSantos-Fernandez/SSNbayes}). It can be installed using:

\begin{example}
install.packages("SSNbayes", dependencies = T)
\end{example}

\noindent Or: 

\begin{example}
remotes::install_github("EdgarSantos-Fernandez/SSNbayes", dependencies = T)          
\end{example}

\pkg{SSNbayes} requires an R version greater than or equal to 3.3.0.
The \pkg{SSNbayes} package extends the models implemented in the \pkg{SSN} package to account for both spatial and temporal dependence using Bayesian inference.
However, detailed spatial, topological, and attribute data are needed to fit these models. The models rely on a .ssn object, which may or may not be created outside of R. When the aim is to fit models to real data, vector editing, information generation, and formatting can be undertaken in ArcGIS version $\geq{9.3.1}$ \citep{esri} using the Spatial Tools for the Analysis for River Systems (STARS) custom toolset \citep{peterson2014stars}. Alternatively, the \pkg{openSTARS} package \citep{openSTARS} can be used to prepare data in raster format. When the pre-processing is complete, both tools create a new directory with the extension .ssn, which contains all of the spatial, topological, and attribute data needed to fit models to stream network data. This includes shapefiles of the stream network, observed locations, and prediction locations (optional). It also contains the response (optional), covariates (optional), and the information needed to generate hydrologic distances and spatial weights between observed and prediction locations. Several unique identifiers are also assigned to observed and prediction locations to denote unique locations (locID) and unique measurements in space and time (pid). If real data are not being used, the createSSN and SimulateOnSSN functions found in \pkg{SSN} can also be used to generate artificial .ssn objects that meet these requirements. In the absence of .ssn object, \pkg{SSNbayes} can still be used to fit models based solely on Euclidean covariance models, which is not the case in \pkg{SSN}.

\subsection{Motivating dataset and application: Stream temperature}
\label{sec:app}

In this section, we introduce a stream temperature dataset collected using in-situ sensors deployed in the Clearwater River Basin, USA  \citep{isaak2018principal}.
The dataset is used to illustrate how the \pkg{SSNbayes} can be used to explore, analyse and draw conclusions from a Bayesian spatio-temporal model.
For completeness, similar analyses were performed on simulated data and the results are presented in the Appendix. 

A spatial stream network object (.ssn) and the observed/prediction datasets are part of the \pkg{SSNbayes} package. 
For reproducibility, we also created a Kaggle notebook containing the example from this section (\url{https://www.kaggle.com/edsans/ssnbayes}).

The .ssn object was generated using the STARS custom toolset \citep{peterson2014stars} and contains 18 observation and 60 prediction locations spaced at 1km intervals along the stream network. 
Hourly temperature recordings were taken at the observation sites but these were averaged to mean daily values for the two-year period spanned by the data set.
The data residing within the .ssn directory is imported into R and converted to an S4 SpatialStreamNetwork object with the following commands:

\begin{example}
path <- system.file("extdata/clearwater.ssn", package = "SSNbayes")
n <- importSSN(path, predpts = "preds", o.write = TRUE)
\end{example}

Next, pair-wise distances are calculated for all observed sites, observed and prediction sites, and prediction sites:

\begin{example}
createDistMat(n, "preds" , o.write=TRUE, amongpred = TRUE)
\end{example}

\noindent We also read in a data frame containing the response and covariates data:

\begin{example}
clear <- readRDS(system.file("extdata/clear_obs.RDS", package = "SSNbayes"))
\end{example}

\begin{figure}[htbp]
  \centering
   \includegraphics[width=4.5in]{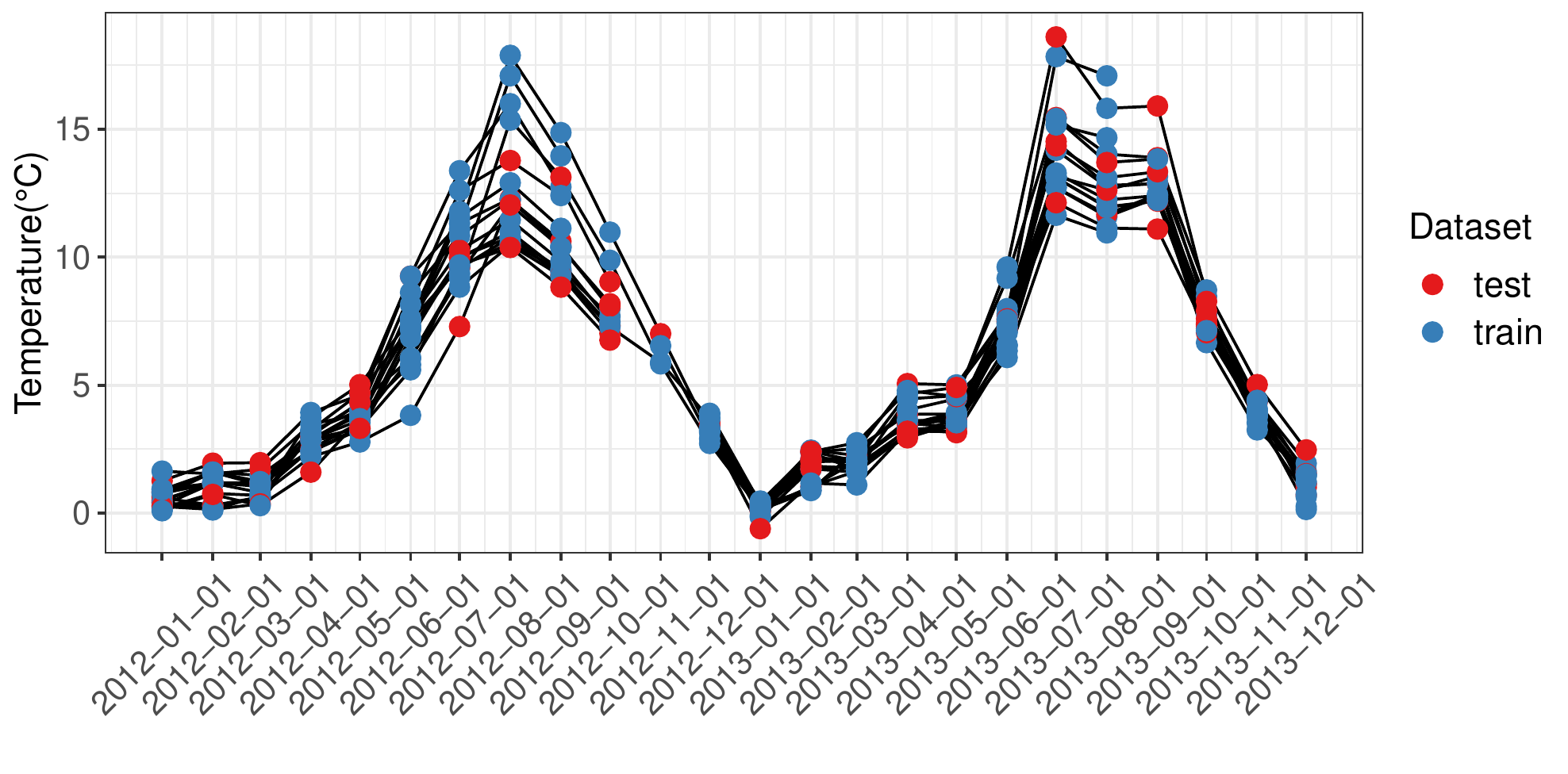}
  \caption{Time series of stream temperatures. Each line represents the time series for a unique observation site. 
  The training dataset is represented by blue points and the testing dataset is shown in red. }
  \label{figure:ts}
\end{figure}

In the data frame \texttt{clear}, the response variable (temp) is the mean daily stream temperature measured at 18 observation sites.
Here we focus on a subsample of longitudinal response data consisting of 24 observations at those sites over two years (Figure \ref{figure:ts}). We randomly split the dataset, with 2/3 used for training the model and 1/3 for testing the out-of-sample prediction accuracy.
This training/testing split was performed once for illustration purposes but for complex datasets we recommend using leave-one-out cross-validation.

Stream temperature is strongly influenced by topography and climate variables \citep{isaak2017norwest}. The following covariates were available for the observation/prediction locations across all the time points: stream slope, elevation, watershed area \citep{isaak2017norwest}, and air temperature  \citep[e.g.][]{bal2014hierarchical}. In addition, we included the first pair of harmonic covariates for the time periods or Fourier terms ($\textrm{sin}_t$ and $\textrm{cos}_t$) \citep{forecast} as covariates.

\subsection{Visualizing stream network data in space and time}

The function \texttt{collapse} extracts line features from a SpatialStreamNetwork object in a format suitable for visualisation using e.g. \pkg{ggplot}. The data frame contains data describing the spatial location of individual stream segments, along with the additive function column.
The function can be used as follows:

\begin{example}
n.df <- collapse(n, par = 'afvArea')
\end{example}

\begin{figure}[htbp]
  \centering
   \includegraphics[width=4.5in]{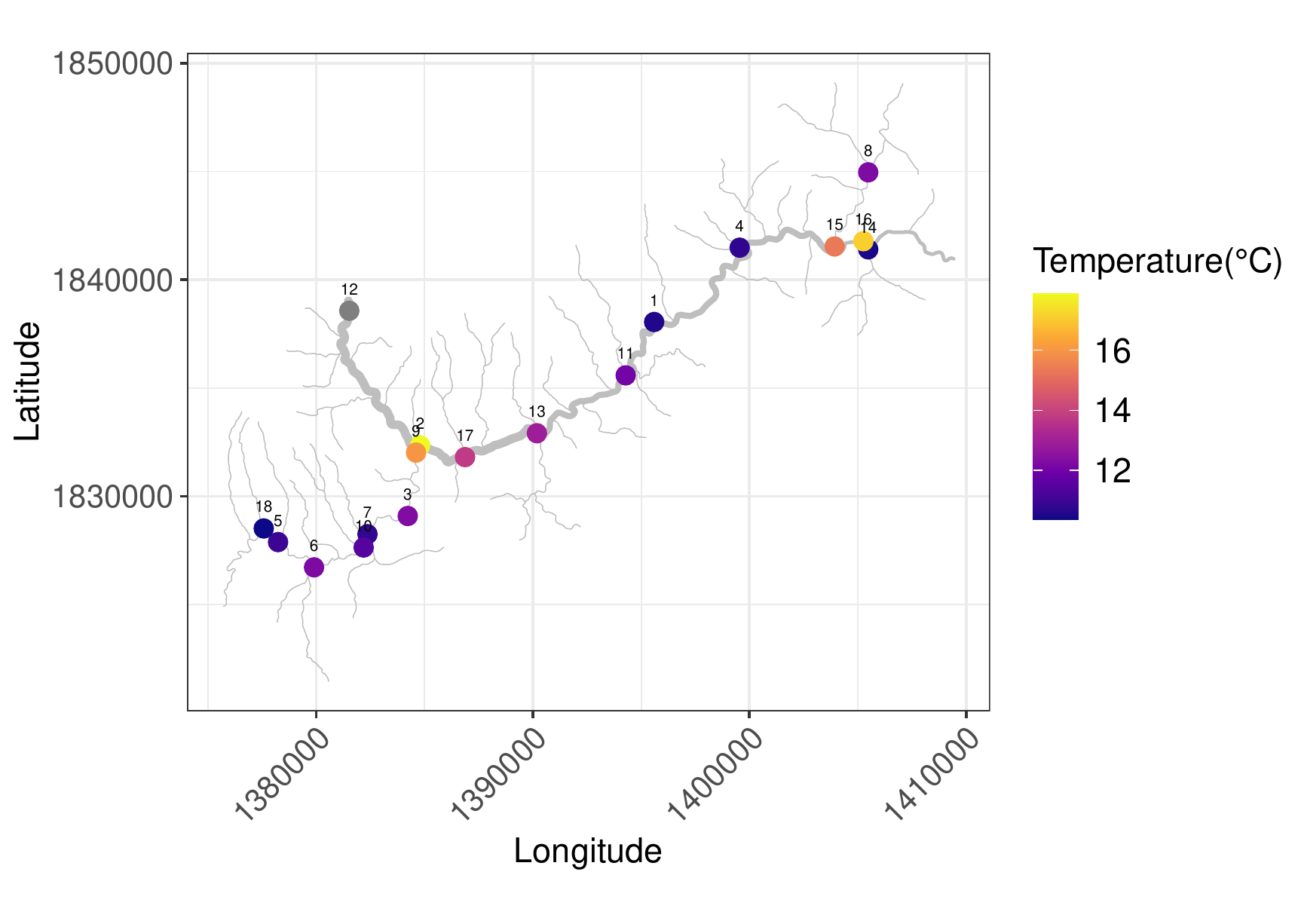}
  \caption{
  Mean daily water temperature in C$^{\circ}$ for 2012-08-01.
  at 18 spatial locations in the Clearwater stream network.
 The labels specify the location identifier ( \texttt{locID}) in the .ssn object.}
  \label{figure:net}
\end{figure}

\noindent The spatial and space-time data can be visualised using \pkg{ggplot2} (Fig~\ref{figure:net}). 

\subsection{Fitting spatio-temporal linear models}

The core function of \pkg{SSNbayes} is \texttt{ssnbayes}. It provides the functionality to fit linear spatio-temporal regression models \citep{santos2021bayesian}.

We specify the following linear regression model using the covariates in the observed dataset: 

\begin{equation}
X_{(t)}^{'}\beta = \beta_0 + \beta_1 * \textrm{SLOPE} + \beta_2 * \textrm{elev}  + \beta_3 * \textrm{cumdrain} + \beta_4 * \textrm{airtemp}_{(t)} +  \beta_5 * \textrm{sin} + \beta_6 * \textrm{cos}. 
\label{eq:lin}
\end{equation}

\noindent We then fit the model to the observed temperature data in the 18 spatial locations with 24 time points in the following way.

\begin{example}
fit_ar <- ssnbayes(formula = temp ~ SLOPE + elev + cumdrainag + air_temp + sin + cos, 
                   data = clear,
                   path = path,
                   space_method = list("use_ssn", "Exponential.taildown"),
                   time_method = list("ar", "date"),
                   iter = 3000,
                   warmup = 1500,
                   chains = 3,
                   net = 2, 
                   addfunccol='afvArea',
                   refresh = max(iter/100,1))
\end{example}

\noindent Running this function takes several minutes and the progress of the sampler is shown during the execution. We have stored the fitted model within \pkg{SSNdata}, which can be accessed using the codes below if the reader wants to skip fitting the model.

\begin{example}
install_github("EdgarSantos-Fernandez/SSNdata")
fit_ar <- readRDS(system.file("extdata//fit_ar.rds", package = "SSNdata"))
\end{example}

\noindent The reader is referred to the Appendix for a second reproducible example using simulated data.

In the function call to \texttt{ssnbayes}, the argument \texttt{formula} describes the regression model and is defined in the same way as other modelling functions such as \texttt{lm} and \texttt{glmssn}. We also pass a data frame using the \texttt{data} argument, which must contain all of the variables specified in the \texttt{formula} argument. This data frame should be in long format, with one row for each unique observation in space and time, which are also defined using \texttt{locID} and \texttt{pid}. In addition, each spatial location must have the same number of temporal observations collected at the exact same times.

The \texttt{space\_method} argument is a list containing information about the spatial modelling component. The first element specifies whether the topological information is stored in a SpatialStreamNetwork object or not  (``use\_ssn'' or ``no\_ssn''), while the second list element specifies which spatial correlation models to use. Options include tail up ("Exponential.tailup", "LinearSill.tailup", "Spherical.tailup"), tail down ("Exponential.taildown", "LinearSill.taildown", "Spherical.taildown") and 
Euclidean ("Exponential.Euclid") models. Certain combinations of spatial covariance matrices are also posible.

If the user specifies use\_ssn as the first element and the second element in the list is missing, then an "Exponential.tailup" model will be used by default. When a tail-up covariance function is specified, an additional column containing the additive function values used to compute the spatial weights must also be specified (e.g. \texttt{addfunccol} ='afvArea'). 
It is possible to have more than one spatial covariance function per family (tail-up, tail-down and Euclidean distance). For instance: \texttt{space\_method = list('use\_ssn', c("Exponential.tailup", "Spherical.taildown"))}.
However, care should be taken in this case to ensure identifiability of the model.
    
The argument \texttt{net} specifies the network identifier when multiple networks are found within the same SpatialStreamNetwork object. Much less information is needed to fit traditional Euclidean covariance models and so a SpatialStreamNetwork object is not needed. Instead, the columns containing the spatial coordinates (e.g. latitude and longitude) must be included as a third element in the list: 
\texttt{space\_method = list("no\_ssn", "Exponential.Euclid", c("lon", "lat"))}. 

The temporal part of the model is defined in a similar fashion using a list \texttt{time\_method = list("method", "date")}. The first element defines the temporal model and options include an autoregressive model, ``ar'', defined in Eq~\ref{eq:case1} or a vector autoregression model, ``var'', defined in Eq~\ref{eq:case2}). The second element is the variable defining the time points, which must be a discrete numeric variable. They should also be spaced at regular intervals, as expected in many time series models. 

In \pkg{SSNbayes} the number of chains (\texttt{chains}), iterations (\texttt{iter}), and burn-in samples (\texttt{warmup}) can be specified. 
By default, \texttt{chains} = 3, \texttt{iter} = 3000, \texttt{warmup} = 1500.
Thinning is also possible using the argument \texttt{thin}. Optionally, the \texttt{seed} parameter can be set to ensure reproducibility.

The \texttt{SSNbayes} package depends on \texttt{Stan}, which does not allow missing values. Therefore, missing values in the response variable are automatically imputed in the \texttt{ssnbayes} function. However, missingness in the covariates is not allowed. Instead, they must be imputed by the user or removed from the dataset before fitting the model. Many options for imputation can be found in \url{https://cran.r-project.org/web/views/MissingData.html}

The output from \texttt{ssnbayes()} is a stanfit object, which contains information about the fitted model and the MCMC chains for the parameters of interest. It can be summarized and visualized using generic functions (\texttt{summary()}, \texttt{plot()}) or functions in the \texttt{ggplot2} package. 

The \texttt{ssnbayes()} function shows the progress of the model fit and will be updated based on the number of samples specified using the \texttt{refresh} argument.  
At every iteration, the inverse of the spatial covariance matrix has to be computed, which takes a substantial amount of time for a 
large number of spatial locations and time points. 
Fitting this dataset using the \texttt{ssnbayes()} function took approximately 10 minutes on a laptop with an Intel Core i7-8650U CPU @ 1.90GHz and 16 Gb of memory.

\subsection{Results}

\begin{figure}[htbp]
  \centering
   \includegraphics[width=6.0in]{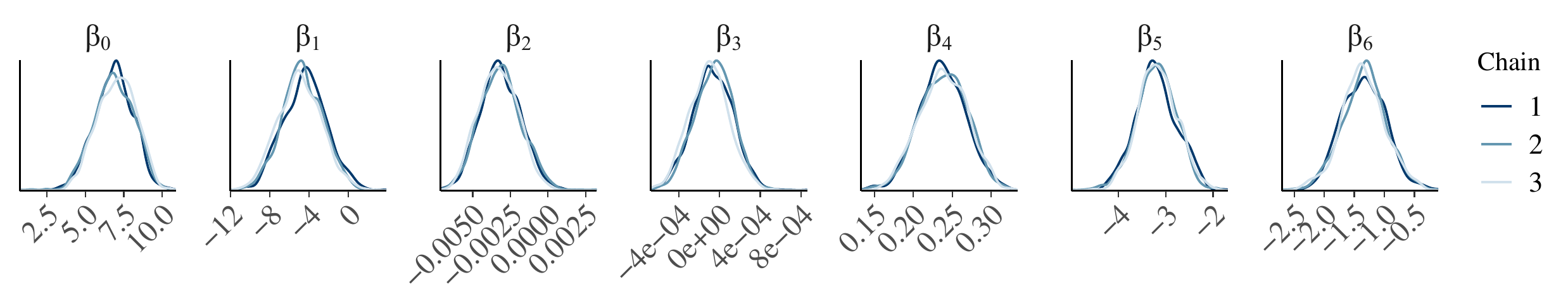}
  \caption{Posterior distributions of the regression coefficients [Intercept ($\beta_0$), stream slope ($\beta_1$), elevation ($\beta_2$), watershed area ($\beta_3$), air temperature ($\beta_4$), $\sin$ ($\beta_5$)and $\cos$ ($\beta_6$)].}
  \label{figure:beta_dens}
\end{figure}

We can visualize the posterior distributions in the parameters of interest.
The regression coefficients from the linear model formulated in Eq~\ref{eq:lin} across three chains are shown in Figure~\ref{figure:beta_dens}.

\begin{example}
mcmc_dens_overlay(
  fit_ar,
  pars = paste0("beta[",1:7,"]"),
  facet_args = list(nrow = 1))
\end{example}

\begin{figure}[htbp]
  \centering
   \includegraphics[width=4.0in]{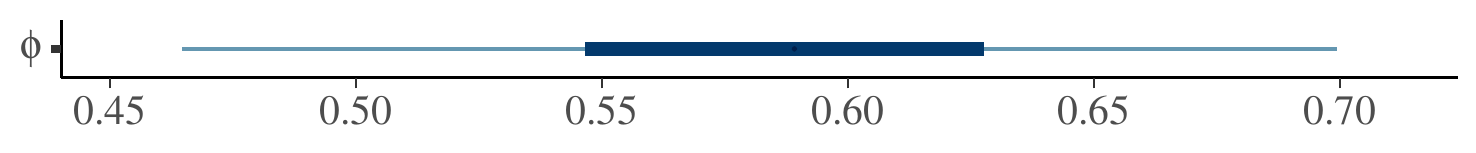}
  \caption{Boxplot of the posterior autoregression parameter, $\phi$.}
  \label{figure:phi_boxplots}
\end{figure}

Apart from the cumdrain area ($\beta_3$), all the estimated regression coefficients for covariates are substantially different from zero.  
The posterior distribution of the  autoregressive parameter ($\phi$), also showed a considerable difference from zero, showing a large amount of temporal dependence (Figure~\ref{figure:phi_boxplots}).

\begin{figure}[htbp]
  \centering
   \includegraphics[width=4.0in]{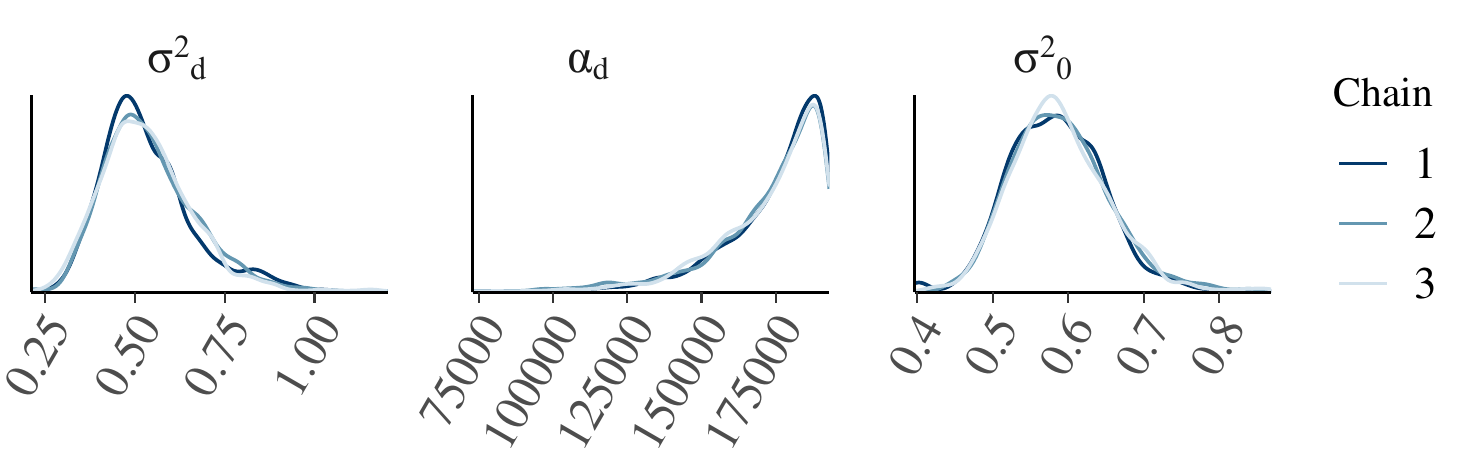}
  \caption{Posterior distributions of the spatial parameters. Units ($\sigma^{2}_d$ and $\sigma^{2}_0$ in C$^{\circ}$, and $\alpha_d$ in meters).}
  \label{figure:var_alpha_nug}
\end{figure}

Figure ~\ref{figure:var_alpha_nug} shows the posterior distributions of the spatial model parameters ($\sigma^2_{TD}$ and $\alpha_{TD}$) and the nugget effect ($\sigma^2_{0}$). Notice that the median of the spatial range $\alpha_{TU}$ is approximately 200,000 m , indicating that spatial autocorrelation exists between locations that are less than 200 km apart.

\begin{example}
mcmc_dens_overlay(
  fit_ar,
  pars = c("var_td", "alpha_td", "var_nug"),
  facet_args = list(nrow = 1))
\end{example}

\begin{figure}[htbp]
  \centering
   \includegraphics[width=6.75in]{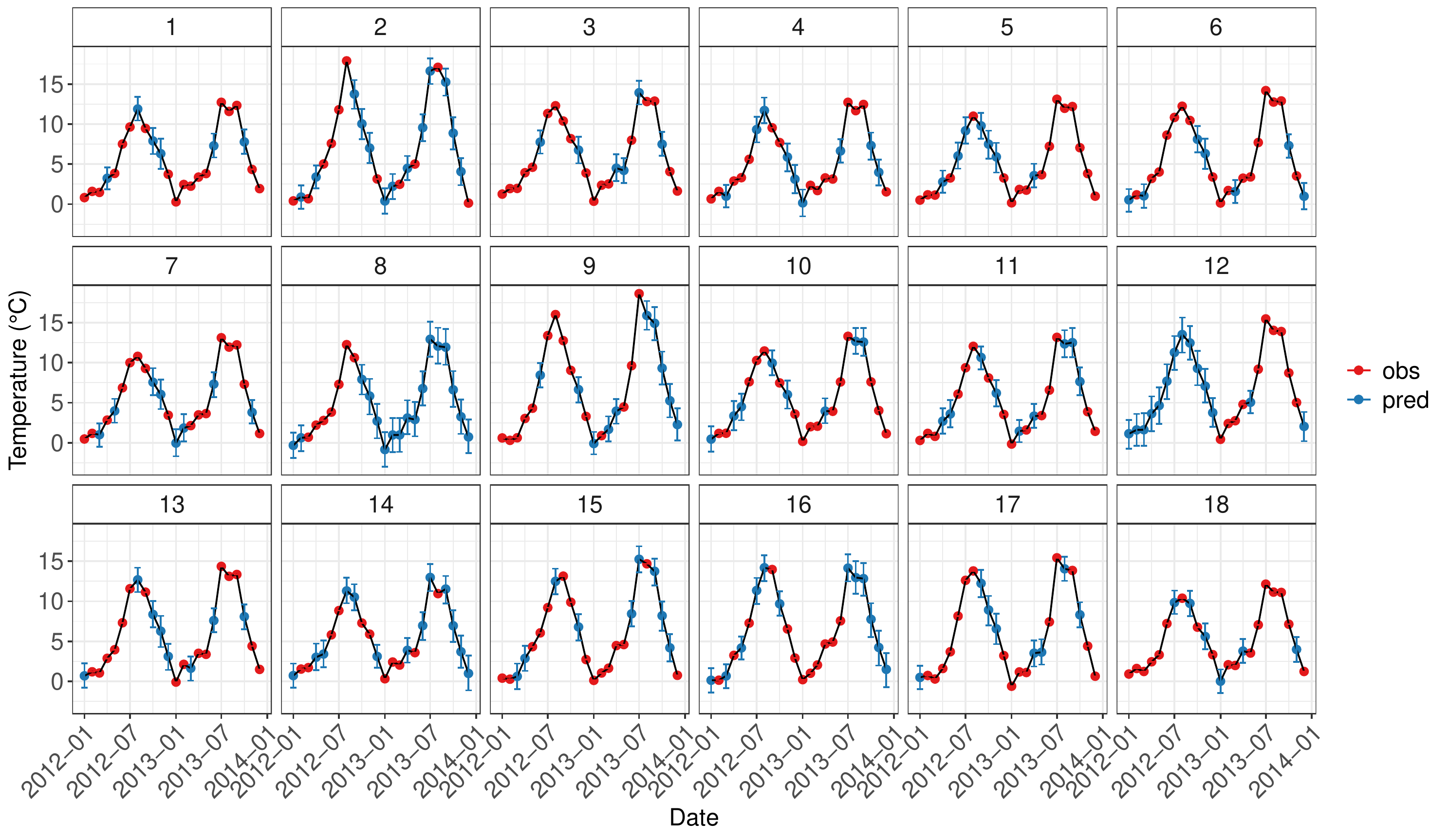}
  \caption{Time series of stream temperatures at 18 spatial locations. 
  The vertical blue bars are the 95\% posterior credible intervals.}
  \label{figure:time_series_obs_pred}
\end{figure}

The time series corresponding to the 18 spatial locations are shown in  Fig~\ref{figure:time_series_obs_pred}.
The observed and predicted points are represented in red and blue respectively.
The model captures the periodic patterns in stream temperatures well, even in locations where most of the observations were missing (e.g. 8 and 12, Fig~\ref{figure:time_series_obs_pred} ).

\begin{figure}[htbp]
  \centering
   \includegraphics[width=6.0in]{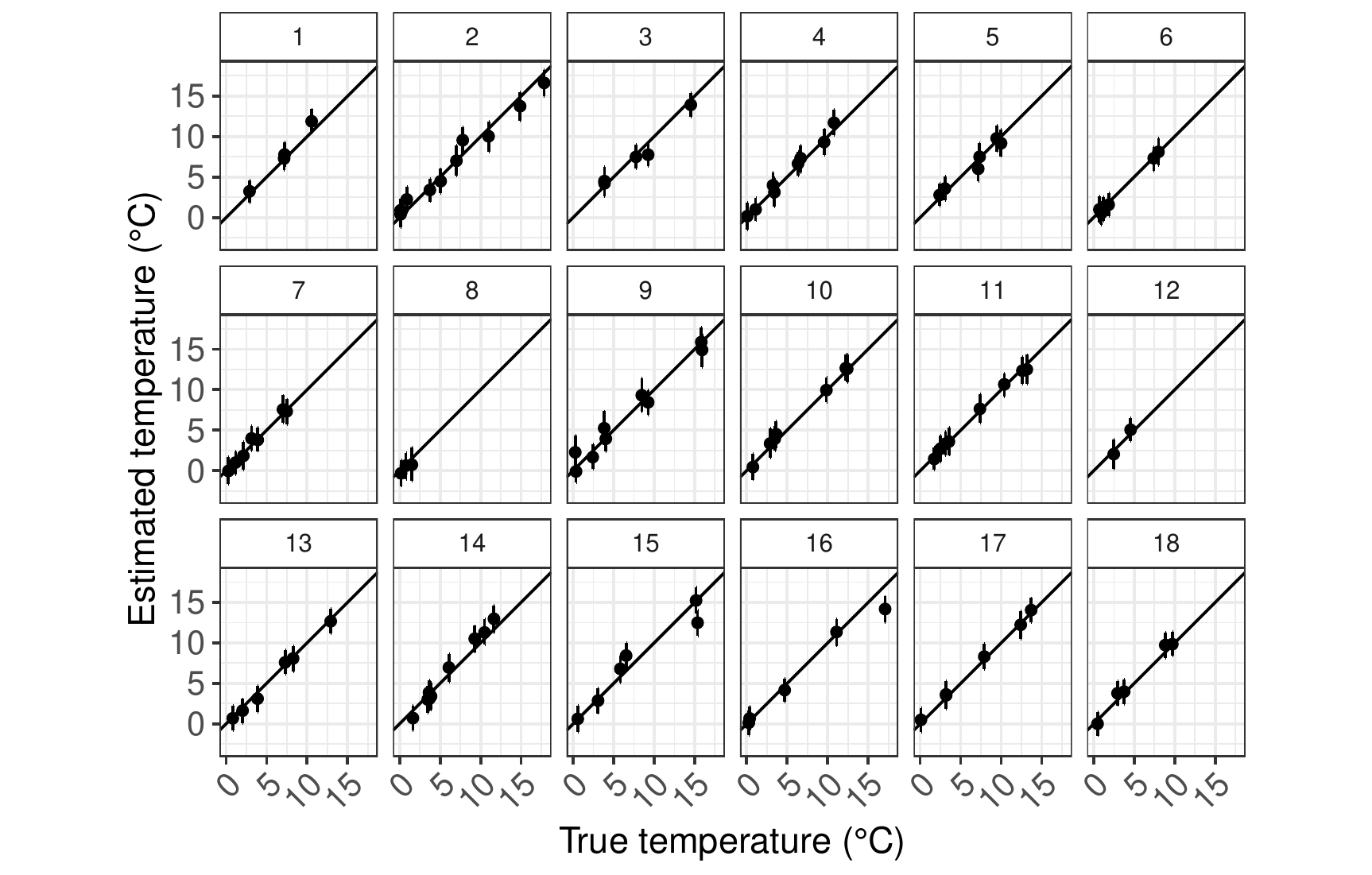}
  \caption{Predicted temperature versus the true latent values at the 18 spatial locations in the test dataset. The vertical bars are the 95\% posterior credible intervals.}
  \label{figure:true_vs_estimated}
\end{figure}

We also compared the predictions produced by the model with the true latent hold-out data. (Fig~\ref{figure:true_vs_estimated}).
If the model predictions were perfect we would expect points to fall on the diagonal line. The results suggest that the Bayesian model produces predictions that are similar to the true latent values. Most of the predictions (96\%) were included within the 
95\% highest density interval, showing appropriate coverage of the predictions.  
The root mean square prediction error (RMSPE) between the true temperature values and the predictions was 0.510 \textdegree{}C, which is small considering the magnitude of the temperature values.

\subsection{Predictions}

Ecological monitoring on stream networks generally produces data at discrete locations, which represent only a small section of the catchment. However, it is often desirable to estimate variables of interest in areas where data have not been collected to create spatially continuous maps \citep{isaak2017norwest}.
In this section, we illustrate how to use the fitted model to predict in unsampled locations using a simple kriging approach.

In our case study, we want to produce temperature predictions at 60 locations generated using a systematic design ($\approx$1km apart). 
The function \texttt{pred\_ssnbayes} produces predictions using information contained in the \texttt{stanfit} object obtained from \texttt{ssnbayes}. The argument  \texttt{nsamples} specifies the number of random samples to select from the posterior distributions and it must be smaller than or equal to the number of iterations  \texttt{iter} specified in \texttt{ssnbayes}.

\begin{example}
pred <- pred_ssnbayes(path = path,
                      obs_data = clear,
                      stanfit = fit_ar, 
                      pred_data = clear_preds,
                      net = 2,
                      nsamples = 100, # number of samples to use from the posterior in the stanfit object 
                      addfunccol = 'afvArea', # variable used for spatial weights
                      locID_pred = locID_pred,
                      chunk_size = 60)

\end{example}

\noindent The observation and prediction data frames (\texttt{data\_obs}, \texttt{data\_pred}, respectively) must be specified and must contain all of the covariates and response variable specified in the ``formula'' argument in \texttt{ssnbayes}.

Generally, producing subsets of predictions on the stream network is more efficient for big datasets, and it can be parallelized. The argument \texttt{chunk\_size} is used to define the size of the subsets. 
\texttt{locID\_pred} also allows the user to define a subset of prediction locations where predictions should be generated, as demonstrated in the example below. Similarly, the argument \texttt{seed} allows the user to set a seed so that the results are reproducible.

\begin{figure}[htbp]
  \centering
   \includegraphics[width=5.0in]{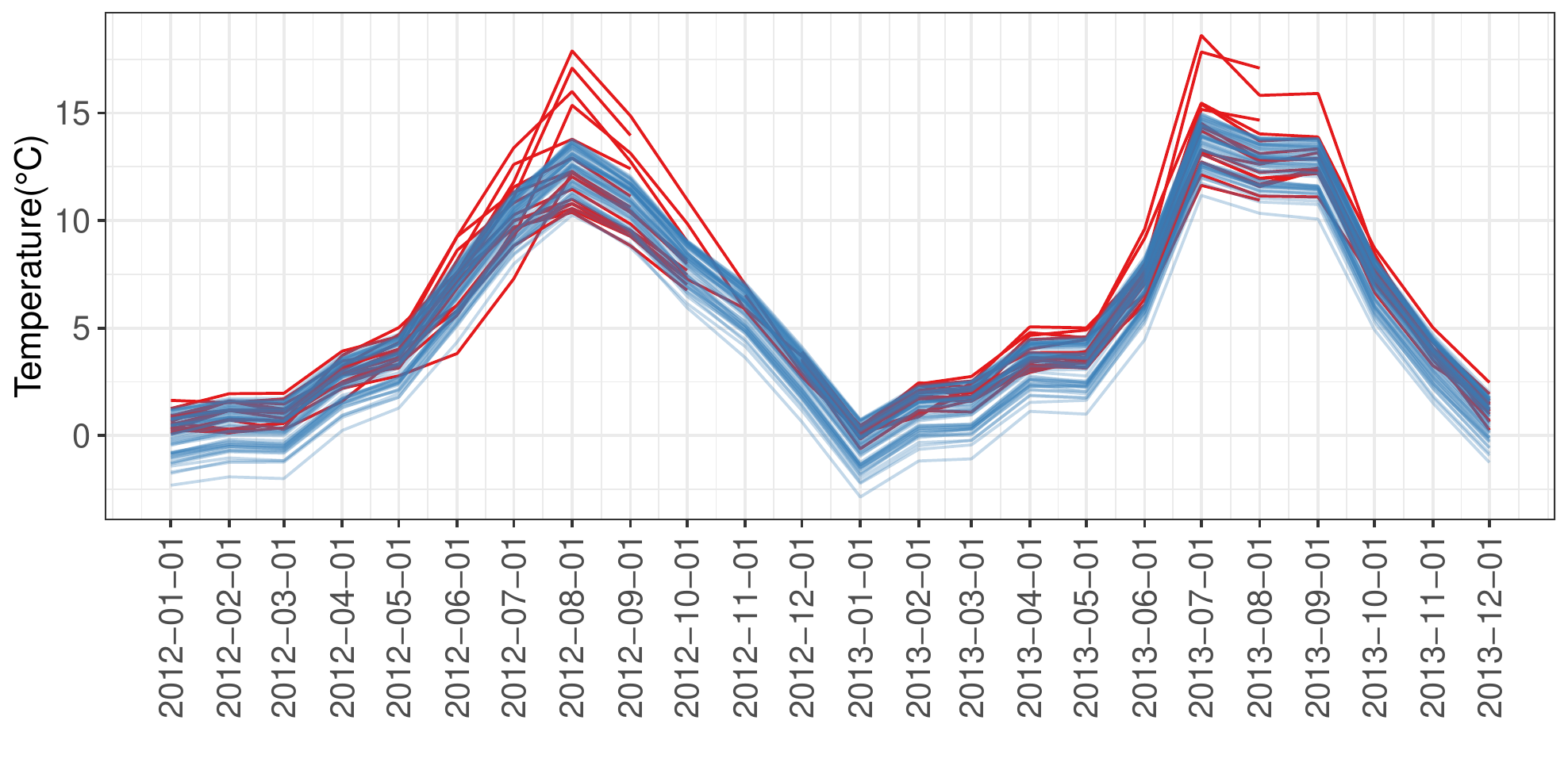}
  \caption{Time series of the predicted (blue lines) and observed temperature (red lines) values. }
  \label{figure:preds_ts}
\end{figure}

\begin{figure}[htbp]
  \centering
   \includegraphics[width=5.75in]{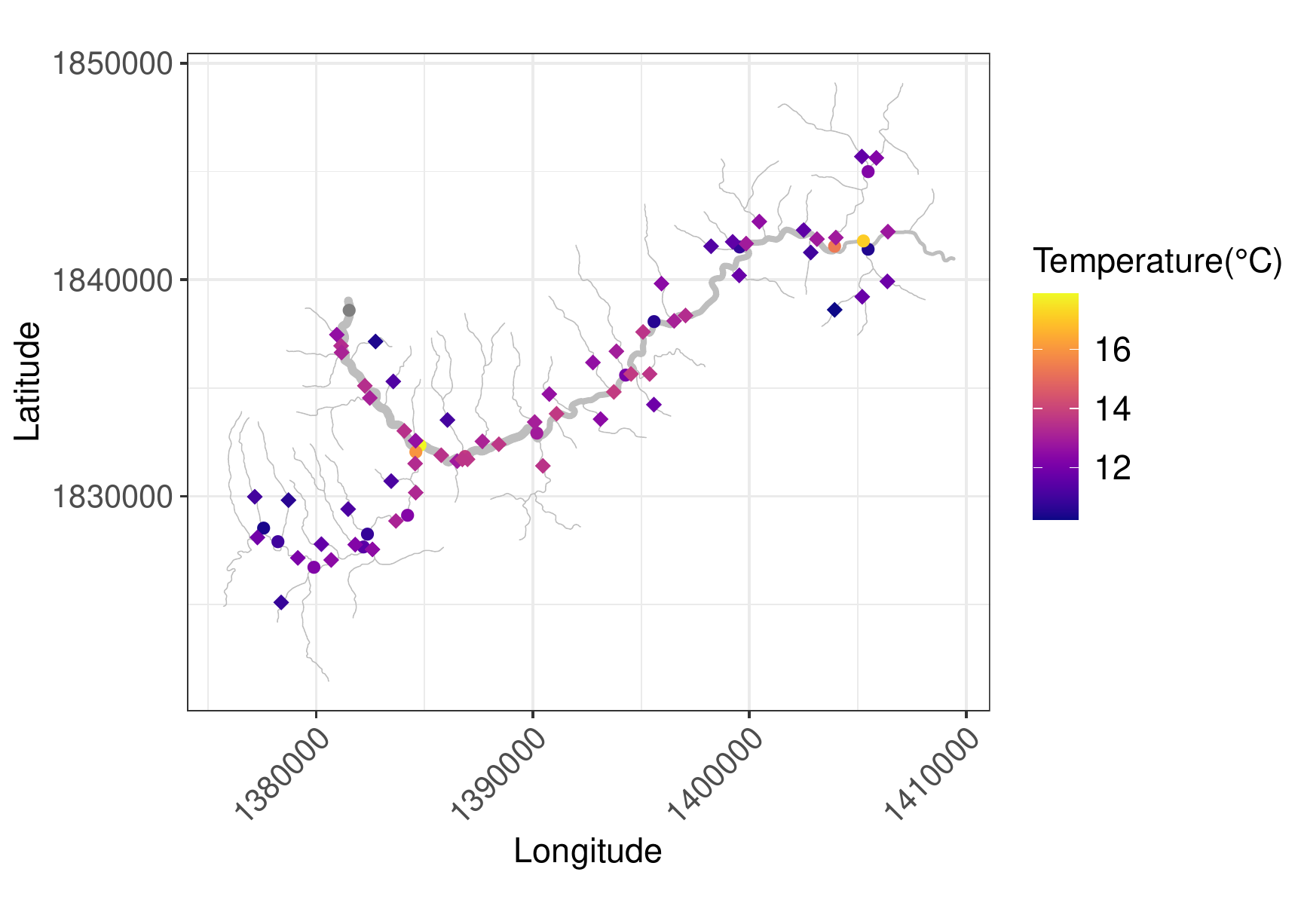}
  \caption{Mean daily stream temperature predictions (diamonds) and observations (circle) in the Clearwater network on 2012-08-01.}
  \label{figure:preds_network}
\end{figure}

Figure ~\ref{figure:preds_ts} shows the predicted time series for the observation and prediction locations. The patterns in the prediction time series captured well the seasonality in the observed data. 
Figure ~\ref{figure:preds_network} visualizes the predictions' posterior temperature means on the stream network. As expected, higher temperature values are obtained in the main stream channel, compared to predictions in small streams which generally are found at higher elevations. 

\subsection*{Network exceedance probability}

\begin{figure}[htbp]
  \centering
   \includegraphics[width=6.95in]{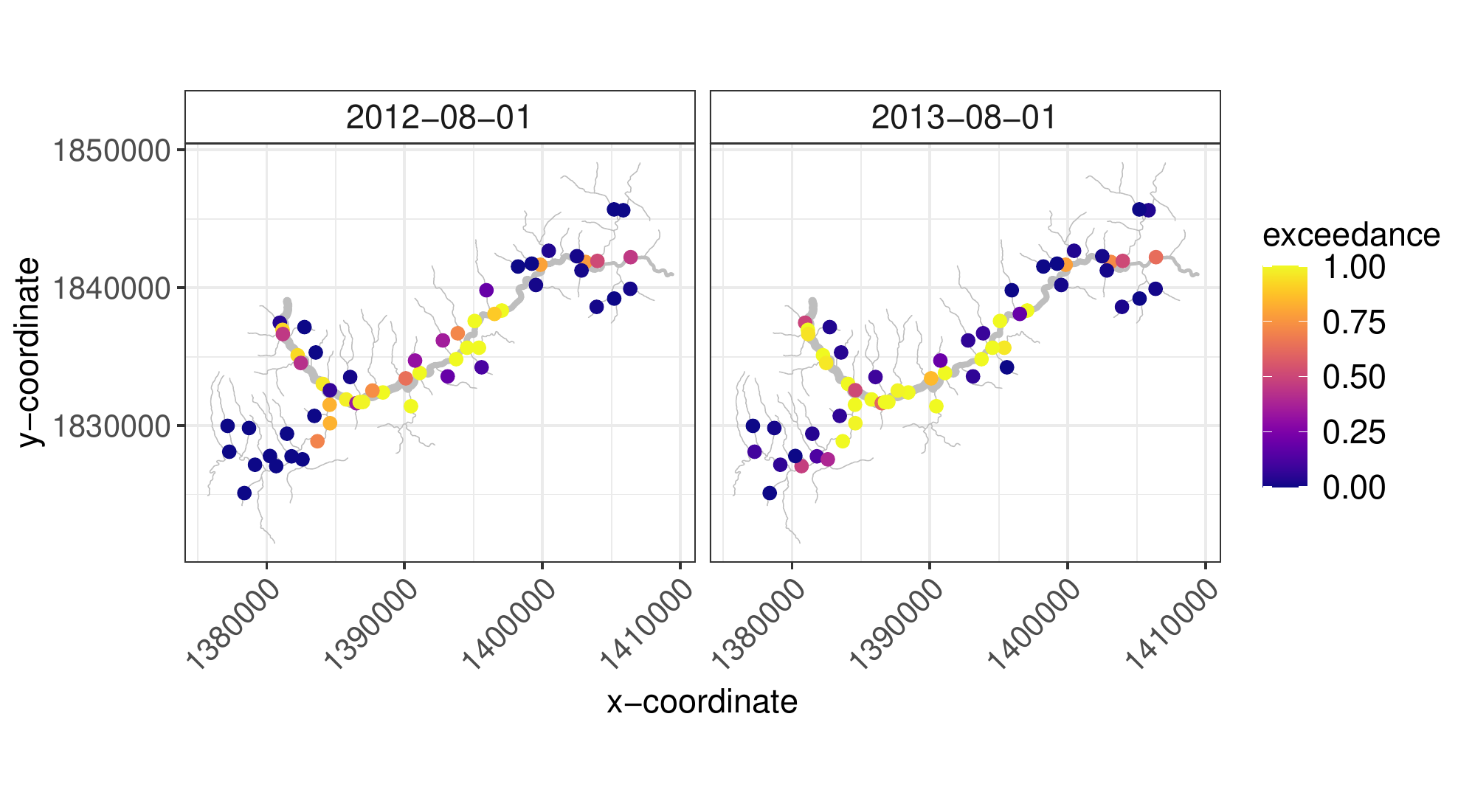}
  \caption{Probabilities that the mean stream temperature will exceed the 13 \textdegree{}C threshold on 2012-08-01 and 2013-08-01.}
  \label{figure:exc}
\end{figure}

It is straightforward to obtain various probabilistic estimates based on the model posterior predictive samples.
In this example, we provide exceedance probabilities based on a critical thermal threshold of 13 \textdegree{}C for bull trout, a cold-water fish species that is sensitive to increased temperatures.
Figure ~\ref{figure:exc} shows the exceedance probabilities for all 60 prediction locations on two dates, obtained from the posterior predictive distributions. Knowledge about when and where biologically relevant thermal thresholds are likely to be exceeded provide critical information for management of threatened and endangered freshwater species \citep{isaak2016slow}.

\begin{example}
ys <- reshape2::melt(pred, id.vars = c('locID0', 'locID', 'date'), value.name ='y')
ys$iter <- gsub("[^0-9.-]", "", ys$variable)
ys$variable <- NULL #$
# network exceedance probability
limit <- 13
ys$exc <- ifelse(ys$y > limit , 1, 0)
ys <- data.frame(ys) 
  dplyr::summarise(sd = sd(y, na.rm=T),
                   y_pred = mean(y, na.rm=T),
                   prop = mean(exc, na.rm=T)) 
  dplyr::arrange(ys, locID)
clear_preds2 <- clear_preds 
\end{example}

\subsection*{Other useful functions}

Other functions that are useful for modelling and computation are \texttt{dist\_wei\_mat()} and \texttt{dist\_wei\_mat\_preds()}. They
produce a list of distance and weight matrices with the following elements:

\begin{enumerate}
    \item e: Euclidean distance matrix containing the distance between locations
    \item D: Downstream hydrologic distance.
    \item H: Total hydrologic distance.
    \item w.matrix: spatial weights for flow connected locations. This matrix is used in the tail-up models. 
    \item flow.con.mat: flow connected matrix. Indicates if two locations in the network are connected by flow.
\end{enumerate}

The \texttt{dist\_wei\_mat()} function produces matrices of the distances and weights between observation locations, with dimensions equal to the number of observation locations ($n_o \times n_o$).
The \texttt{dist\_wei\_mat\_preds()} function produces the same information for observed and prediction locations, with  $n_o n_p \times n_o n_p$ dimensions.
Details and detailed descriptions of the computation of these matrices can be found in \citet{peterson2010mixed}, \citet{hoef2014ssn}, and \citet{santos2021bayesian}.

\section{Discussion and conclusions}
\label{sec:dis}

The growth of stream sensor arrays in which repeat observations are taken at multiple sites requires models capable of accounting for 
spatial and temporal autocorrelation to stream network data.    
However, there are only a limited number of computational methods and software packages that were designed to account for the unique spatial dependence found in streams data (e.g. the R packages \pkg{SSN} and \pkg{smnet}). 
This package extends the models implemented in \pkg{SSN} by accounting for temporal dependence using Bayesian inference, which offers several benefits.
Enhanced features from this package are the computation of network exceedance probabilities and other benefits from the use of a Bayesian framework, including the ability to incorporate prior information, probabilistic estimates and proportion of degraded habitat.   

We have tested the performance of \pkg{SSNbayes} in multiple scenarios with simulated and real data and we have found that the parameters are well estimated and the predictions are accurate in terms of RMSPE. 
We have also validated the results from a wide range of spatial model combinations to those obtained using \pkg{SSN} based on simulated data.  
Spatial and spatio-temporal models tend to be slow and computationally intensive, which becomes more challenging within a Bayesian modelling framework. This can become computationally prohibitive when the number of spatial locations is large because the spatial covariance matrix must be iteratively inverted. We are currently researching alternative methods that will scale well for models implemented within \pkg{SSNbayes}.

Future implementations will incorporate other modelling variations.
Two of them are: (I) expressing $\phi_s$ as a linear combination of covariates such as elevation and watershed, and (II)
using a 2-Nearest Neigbours (2-NN) method, where the off-diagonal elements of $\Phi$ are different from zero in the two closest, allowing temporal dependence to be established between neighbouring spatial locations connected by flow \citep{santos2021bayesian}. 
However, there are numerous other  space-time covariance structures that could be implemented for stream network data, which allow more modelling flexibility. For example, this implementation is based on a vector autoregression structure, but other models such as moving averages and ARIMA could also be considered. In addition, we currently assume that the response variable is normally distributed, but other regression models could be implemented by modifying the likelihood function in \texttt{ssnbayes}. 
We are also actively working in the development and implementation of models that account for measurement errors and anomalies in the data.
The R package is under constant development and new features and implementations are on their way.

\section*{Acknowledgement}

This research was supported by the Australian Research Council (ARC) Linkage Project ``Revolutionising water-quality monitoring in the information age'' (ID: LP180101151) and the Centre of Excellence for Mathematical and Statistical Frontiers (ACEMS). 
JMM was supported by an Australian Research Council Discovery Project (DP200101263).
We thank Dona Horan for the creation of the spatial stream network (SSN) object.
Data analysis and computations were undertaken using the packages \textsf{rstan} \citep{rstan}. Data visualizations were made with the packages \textsf{tidyverse} \citep{tidyverse} and \textsf{bayesplot} \citep{bayesplot}.

\bibliography{ref}

\address{Edgar Santos-Fernandez\\
  School of Mathematical Sciences. Queensland University of Technology\\
  Australian Research Council Centre of Excellence for Mathematical and Statistical Frontiers (ACEMS)\\
  Y Block, Floor 8, Gardens Point Campus. GPO Box 2434. Brisbane, QLD 4001. \\
  Australia\\
  (0000-0001-5962-5417)\\
  \email{santosfe@qut.edu.au}}

\address{Jay M. Ver Hoef\\
Marine Mammal Laboratory\\
NOAA-NMFS Alaska Fisheries Science Center
Seattle, WA and Fairbanks, AK, USA\\
(0000-0003-4302-6895)\\
\email{jay.verhoef@noaa.gov}}

\address{James McGree \\
School of Mathematical Sciences. \\
Queensland University of Technology\\
(0000-0003-2997-8929)\\
\email{james.mcgree@qut.edu.au}}

\address{Daniel J. Isaak\\
  Rocky Mountain Research Station. US Forest Service \\
  \email{disaak@fs.fed.us}}

\address{Kerrie Mengersen\\
  School of Mathematical Sciences. Queensland University of Technology\\
  Australian Research Council Centre of Excellence for Mathematical and Statistical Frontiers (ACEMS)\\
  Y Block, Floor 8, Gardens Point Campus. GPO Box 2434. Brisbane, QLD 4001. \\
  Australia\\
  (0000-0001-8625-9168)\\
  \email{k.mengersen@qut.edu.au}}

\address{Erin E. Peterson\\
  School of Mathematical Sciences. Queensland University of Technology\\
  Australian Research Council Centre of Excellence for Mathematical and Statistical Frontiers (ACEMS)\\
  Y Block, Floor 8, Gardens Point Campus. GPO Box 2434. Brisbane, QLD 4001. \\
  Australia\\
  (0000-0003-2992-0372)\\
  \email{erin@peterson-consulting.com}}

\clearpage

\appendix
\label{sec:appndx} 

\subsection*{Using simulated data}

We start by generating some spatial data with the \pkg{SSN} package using a systematic design for the locations of the observations and predictions:

\begin{example}
seed <- 202008
set.seed(seed)
path <- "./sim.ssn"

n <- createSSN(n = c(150),  # segments
               obsDesign = systematicDesign(3),
               predDesign = systematicDesign(0.3), 
               importToR = TRUE,
               path = path, # path where the sns object is saved
               treeFunction = iterativeTreeLayout)
							
(points <- nrow(getSSNdata.frame(n, "Obs"))) # numb of observation locations
nrow(getSSNdata.frame(n, "preds")) # numb of prediction locations
\end{example}

This produced the SSN object with 50 observation locations, which is our training dataset. We also generated for testing 499 prediction locations. 
Consider that we want to model a response variable (stream temperature). 
The aim is the to predict this response variable in the testing dataset borrowing information across space and time and using covariates. This prediction can be in some locations or across the whole network.

We then generate the distance/weight matrices within and between observation and predictions.
We then need to simulate some data using some covariates, regression coefficients and 
a covariance structure.

\begin{example}
createDistMat(n, o.write=TRUE) # creates the distance/weight matrices
rawDFobs <- getSSNdata.frame(n, "Obs")

# generating 3 continous covariates
rawDFobs[,"X1"] <- rnorm(length(rawDFobs[,1]))
rawDFobs[,"X2"] <- rnorm(length(rawDFobs[,1]))
rawDFobs[,"X3"] <- rnorm(length(rawDFobs[,1]))

n <- putSSNdata.frame(rawDFobs,n, Name = 'Obs')

set.seed(seed)
# We simulate some data using the covariates, regression coefficients and 
# a taildown exponential covariance structure
sim.out <- SimulateOnSSN( n, ObsSimDF = rawDFobs,
                          formula = ~ X1 + X2 + X3,  coefficients = c(10, 1, 0, -1),
                          CorModels = c("Exponential.taildown"), use.nugget = TRUE,
                          CorParms = c(3, 10, .1), addfunccol = "addfunccol")

sim.ssn <- sim.out$ssn.object #$ NB: so that is not taken as an Eq by Latex

\end{example}

We now need to generate some time series with AR(1) error structure:
\begin{example}
obs_data <- getSSNdata.frame(sim.ssn, "Obs")
t <- 10 # number of days
obs_data <- do.call("rbind", replicate(t, obs_data, simplify = FALSE))# replicating the obs_data df
obs_data$date <- rep(1:t, each = (nrow(obs_data)/t))   

set.seed(seed)
phi <- 0.8 
ar1 <- nlme::corAR1(form = ~ unique(obs_data$date), value = phi) 
AR1 <- Initialize(ar1, data = data.frame(unique(obs_data$date)))

epsilon <- t(chol(corMatrix(AR1))) 

epsilon <- rep(epsilon, each = length(unique(obs_data$locID)) ) + 
  rnorm(length(epsilon)*length(unique(obs_data$locID)), 0, 0.25) # for the 10 dates


obs_data$epsilon <- epsilon
obs_data$y <- obs_data$Sim_Values + obs_data$epsilon # adding the error term
obs_data$pid <- rep(1:nrow(obs_data)) # generating a new pid
\end{example}

Visualizing the time series on the 50 locations:
\begin{example}
ggplot(obs_data) + 
    geom_line(aes(x = date, y = y, group = locID)) +
    theme_bw()
\end{example}

\begin{figure}[htbp]
  \centering
   \includegraphics[width=4.0in]{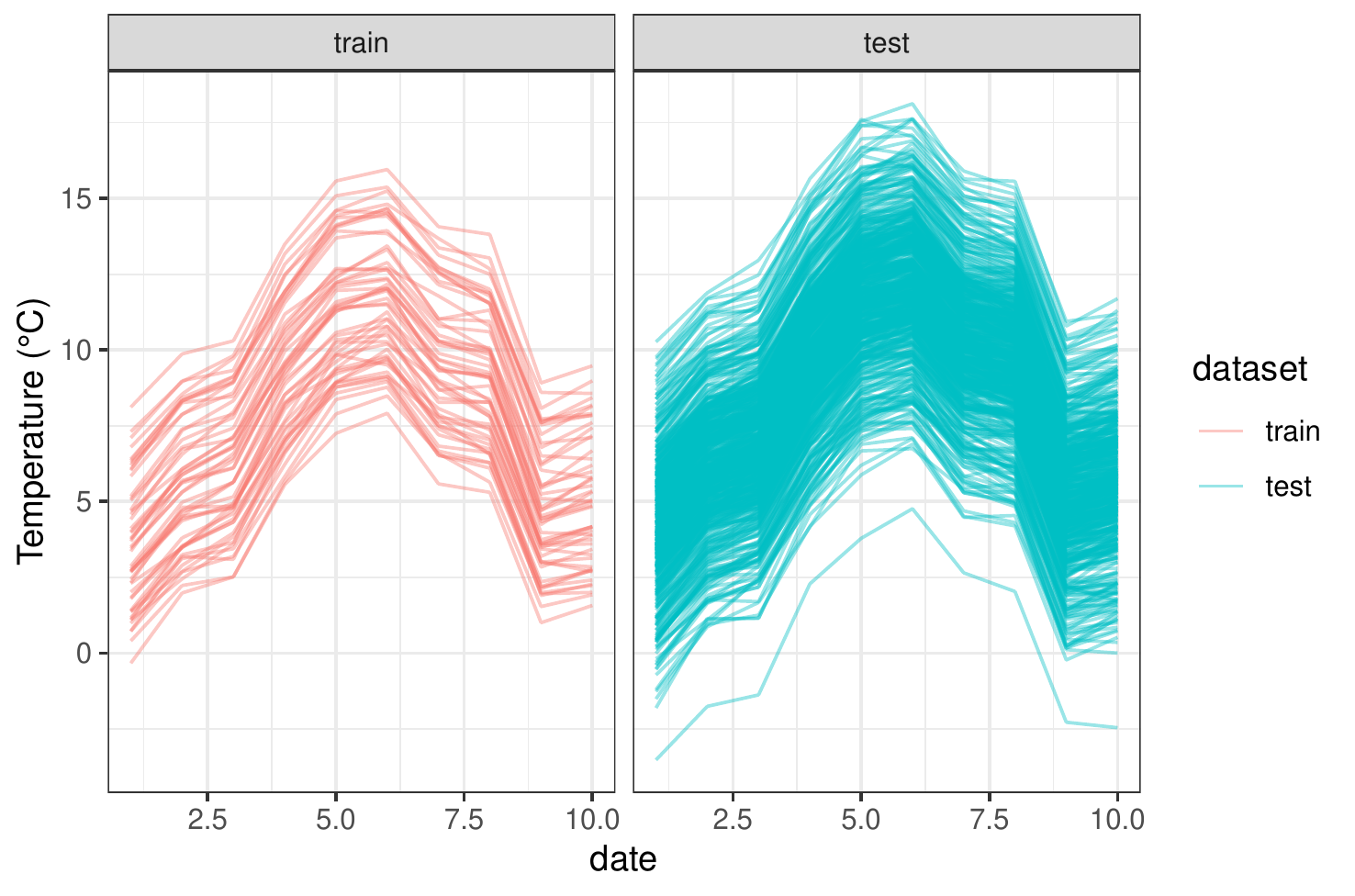}
  \caption{Evolution of the stream temperature time series in the training and testings datasets.}
  \label{figure:1time_series}
\end{figure}

Figure~\ref{figure:1time_series} shows the stream temperature time series in the training (observations) and testings datasets (predictions).

Let us set 30\% of the observations per date to missing to assess how well we retrieve the latent temperature values. 
This is the first alternative for prediction within \pkg{SSNbayes}. 

\begin{example}
set.seed(seed)
locs <- obs_data 
  do(sample_n(., round(points*0.3), replace = F)) 

#obs backup
obs_data$y_backup <- obs_data$y # backing up the temeprature values
obs_data[obs_data$pid 

obs_data_coord <- data.frame(n@obspoints@SSNPoints[[1]]@point.coords) # extracting the cooordinates
obs_data_coord$locID <- factor(1:nrow(obs_data_coord))
obs_data_coord$locID <- as.numeric(as.character(obs_data_coord$locID))
obs_data$locID <- as.numeric(as.character(obs_data$locID))
obs_data <- obs_data 
obs_data$point <- 'Obs' #$
\end{example}

Let us visualize the network with the time series of observed temperature values.
We collapsing the SSN object to extract the network structure.
The facets represent the date (1-10).
In each date there are 15 gray dots which are observations that we set to missing to assess the model predictive accuracy.

\begin{example}
nets <- SSNbayes::collapse(n, par = 'addfunccol' )
nets$afv_cat <- cut(nets$computed_afv, 
                                 breaks = seq(min(nets$computed_afv),
                                              max(nets$computed_afv),
                                              length.out=6),
                                 labels = 1:5,
                                 include.lowest = T)

ggplot(nets) + 
  geom_path(aes(X1, X2, group = slot, size = afv_cat), lineend = 'round', linejoin = 'round', col = 'lightblue')+
  geom_point(data = dplyr::filter(obs_data, date 
  scale_size_manual(values = seq(0.2,2,length.out = 5))+
  facet_wrap(~date_num, nrow = 2)+
  scale_color_viridis(option = 'C')+
  scale_shape_manual(values = c(16,15))+
  xlab("x-coordinate") +
  ylab("y-coordinate")+
  theme_bw()
\end{example}

\begin{figure}[htbp]
  \centering
   \includegraphics[width=6.0in]{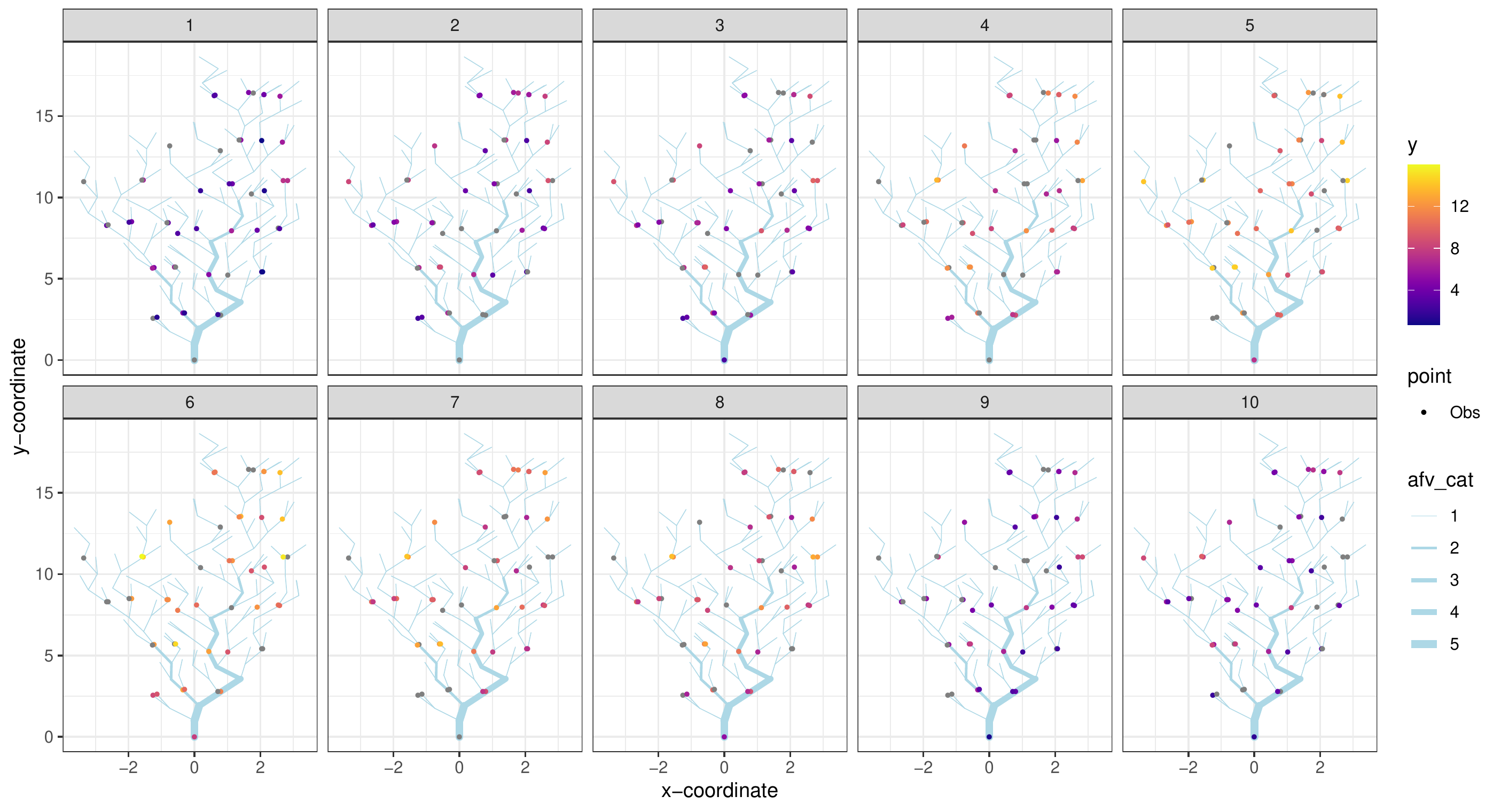}
  \caption{Evolution of the stream temperature time series in the training and testings datasets.}
  \label{figure:2network_time_obs}
\end{figure}

\pagebreak

We then fit a tail-down model with AR(1) error structure using the \texttt{ssnbayes} function:

\begin{example}
pred_data_coord <- data.frame(n@predpoints@SSNPoints[[1]]@point.coords)
pred_data_coord$locID <- factor(1:nrow(pred_data_coord))
pred_data_coord$locID <- as.numeric(as.character(pred_data_coord$locID))
pred_data_coord$locID <- length(unique(obs_data$locID)) + pred_data_coord$locID
pred_data$locID <- as.numeric(as.character(pred_data$locID))
pred_data <- pred_data 
pred_data$point <- 'pred' #$


fit_td <- ssnbayes(formula = y ~ X1 + X2 + X3, 
                   data = obs_data,
                   path = path,
                   time_method = list("ar", "date"), # temporal model to use
                   space_method = list('use_ssn', c("Exponential.taildown")), # spatial model to use
                   iter = 4000,
                   warmup = 2000,
                   chains = 3,
                   addfunccol = 'addfunccol',
                   loglik = T)
\end{example}

\begin{figure}[htbp]
  \centering
   \includegraphics[width=5.0in]{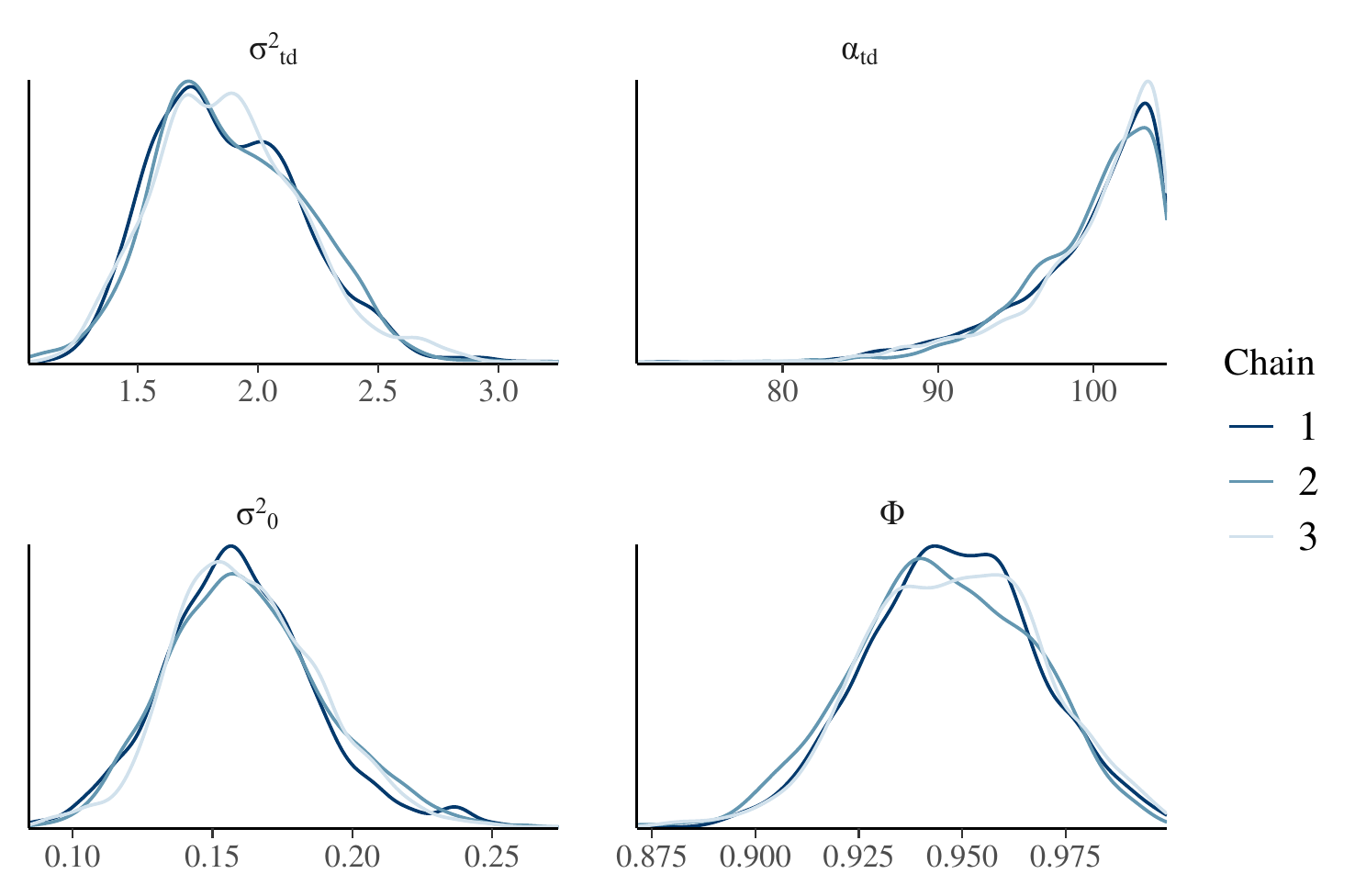}
  \caption{Posterior densities of the partial sill ($\sigma^2_{TU}$), nugget effect ($\sigma^2_{0}$), range ($\alpha$) and $\phi$ }
  \label{figure:3densities_pars}
\end{figure}

\begin{figure}[htbp]
  \centering
   \includegraphics[width=5.0in]{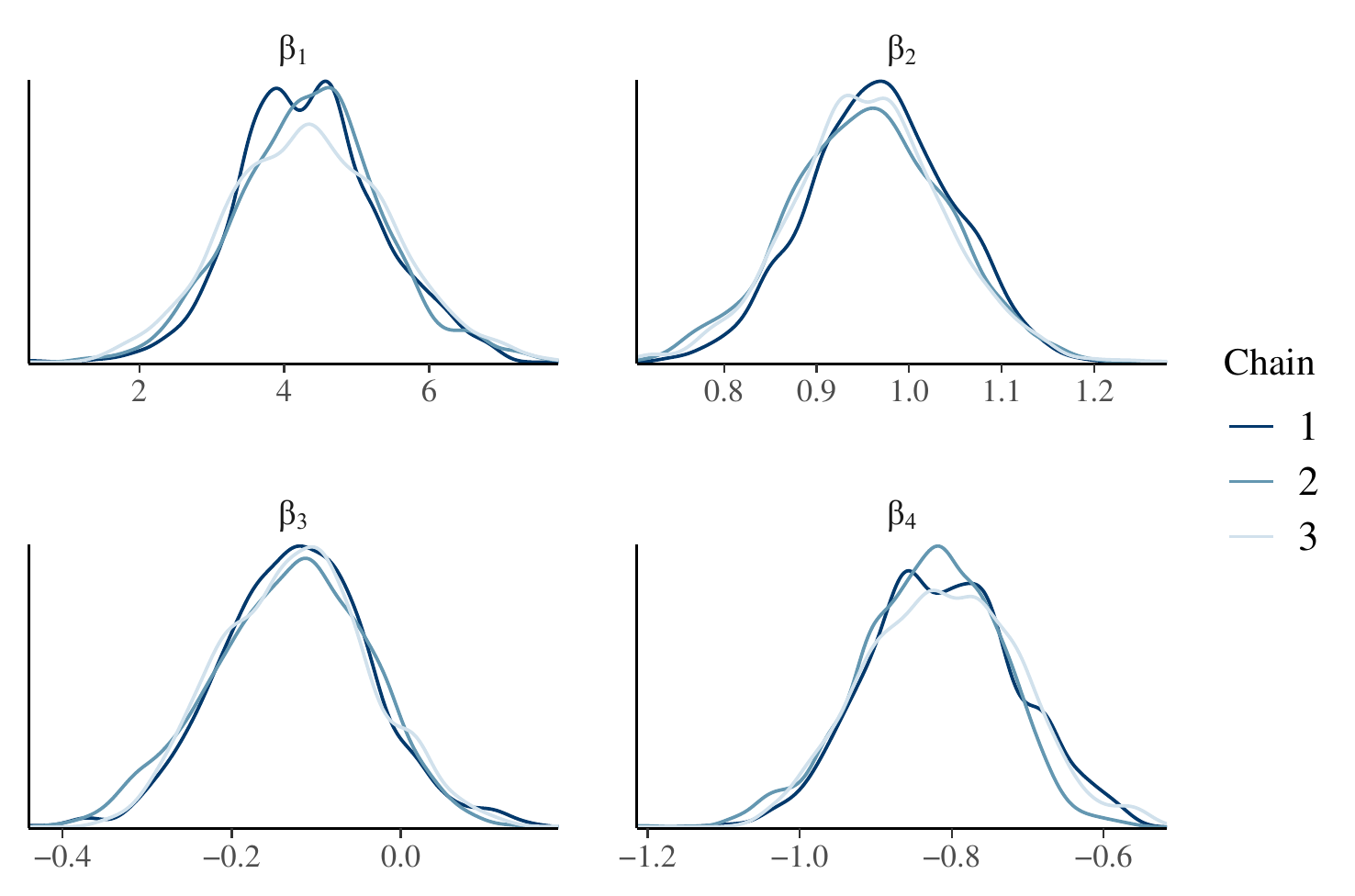}
  \caption{Posterior densities of the regression coefficients ($\beta$). }
  \label{figure:3densities_beta}
\end{figure}

One of the main benefits of this Bayesian approach is that the model produce probabilistic estimates.
Figures ~\ref{figure:3densities_pars} and  ~\ref{figure:3densities_beta} show the posterior distributions of the four parameters in the spatio-temporal model ($\sigma^2_{TU}$, $\sigma^2_{0}$), $\alpha$ and $\phi$ ) and of the regression coefficients (intercept and slopes).  
The trace plots for of these parameters can be found in the \ref{sec:appndx}.

\begin{figure}[htbp]
  \centering
   \includegraphics[width=5.0in]{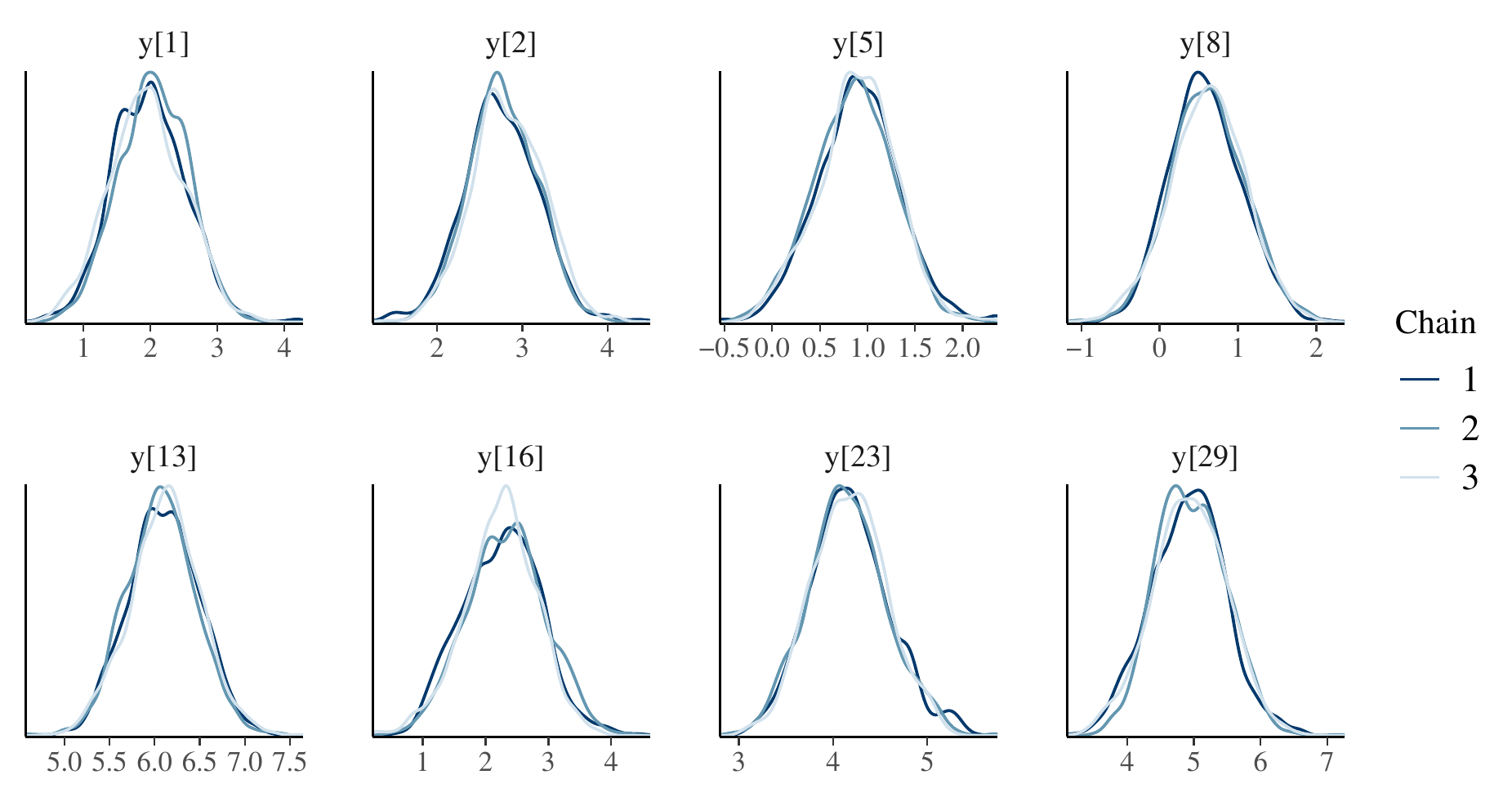}
  \caption{Posterior densities of the predicted temperature ($y$). }
  \label{figure:3densities_y}
\end{figure}

\begin{figure}[htbp]
  \centering
   \includegraphics[width=5.0in]{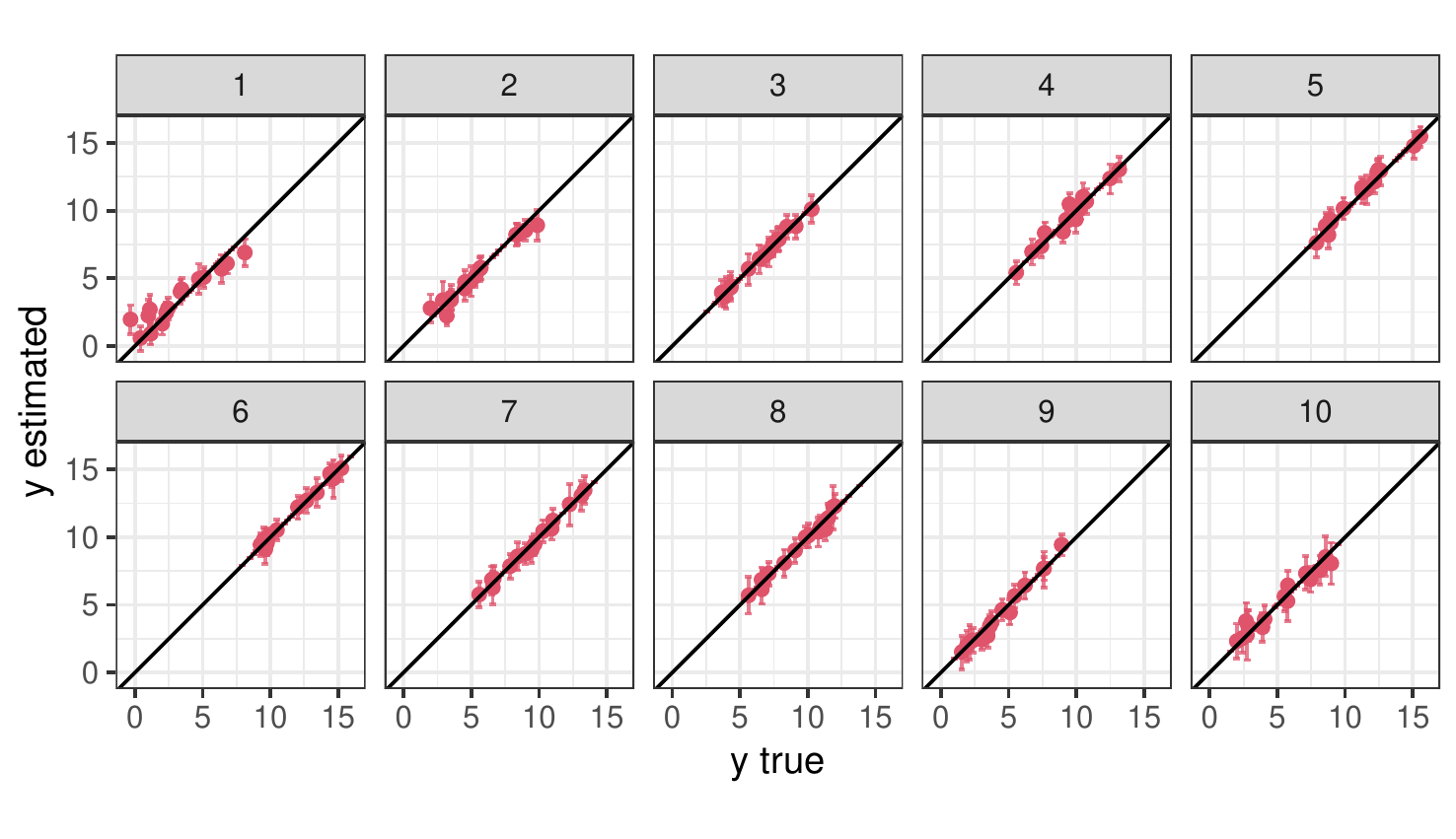}
  \caption{Predicted temperature versus the true latent values. 
  The error bars represent the 95\% highest density interval}
  \label{figure:4y_true_vs_pred}
\end{figure}

We then assess how good are the predictions of the missing temperature values comparing the true latent versus the predicted temperature values.
If the model prediction was perfect we would expect points falling in the diagonal line.  
From Figure ~\ref{figure:4y_true_vs_pred} we see that the Bayesian model produces estimates very similar to the latent values in the 10 day. Most of the prediction points (96\%), are included within the 
 95\% highest density interval showing goood coverage of the estimates.  
The RMSE is 0.245 which is small compared to the magnitude of the response variable with mean 7.621 and sd = 3.489 degrees.

\begin{example}
fits <- fit_td
class(fits) <- c("stanfit")

stats_td <- summary(fits)
stats_td <- stats_td$summary


## Create plots of chain results for seven reg coef
mcmc_dens_overlay(
  fits, # 
  pars = paste0("beta[",1:3,"]"),
  facet_args = list(nrow = 1)) 

## Plot the distribution of phi
mcmc_intervals(
  fit,
  pars = paste0("phi"),
  point_size = .1,
  prob_outer = 0.95
) 


## Plot the nugget effect, partial sill and range
## parameter distributions for the tail-down model
mcmc_dens_overlay(
  fit,
  pars = c(
    "var_td",
    "alpha_td",
    "var_nug"),
  facet_args = list(nrow = 1)
)

# How good are the predictions?

ypred <- data.frame(stats_td[grep("y\\[", row.names(stats_td)),])
ypred$ytrue <- obs_data$y_backup #
ypred$date <- rep(1:t, each = 50)
ypred$dataset <- ifelse(ypred$sd == 0, 'obs', 'pred')
ypred$td_exp <- ypred$mean

 filter(ypred, dataset == 'pred') 
  geom_errorbar(data = ypred, aes(x=ytrue, ymin=X97.5., ymax=X2.5.), col = 2, width=0.5, size=0.5, alpha = 0.75) +
  geom_point(aes(x = ytrue , y = td_exp), col = 2)+
  geom_abline(intercept = 0, slope = 1)+
  facet_wrap(~date, nrow = 2) +  coord_fixed() +  
  xlab('y true') + ylab('y estimated') +  theme_bw()

(rmse <- sqrt(mean( ( (ypred$ytrue) - ypred$td_exp)^2)))

# coverage
ypred$cov <- ifelse(ypred$ytrue > ypred$X2.5. & ypred$ytrue<ypred$X97.5,1,0)
filter(ypred, dataset == 'pred') 
  group_by(dataset) 
  dplyr::summarize(mean(cov))
\end{example}

\end{article}

\end{document}